\documentclass[trackchanges]{aastex701}

\usepackage{makecell}
\usepackage{graphicx}
\usepackage{subcaption}
\usepackage{caption} 
\usepackage{booktabs}
\usepackage{amssymb}
\usepackage{amsmath}
\usepackage{float}
\usepackage{xtab}
\usepackage{orcidlink}
\usepackage{adjustbox}

\begin{document}

\title{Anisotropy of Satellite Galaxies-II: Spatial Segregation as the Origin of Anisotropic Satellite Quenching in Haloes and Beyond}

\correspondingauthor{Weiguang Cui, Yun Chen}
\email{weiguang.cui@uam.es, chenyun@bao.ac.cn}

\author[orcid=0000-0002-4518-6035]{Zhuoming Zhang}
\affiliation{National Astronomical Observatories, Chinese Academy of Sciences, Beijing 100101, China}
\affiliation{College of Astronomy and Space Sciences, University of Chinese Academy of Sciences, Beijing, 100049, China}
\email{} 

\author[orcid=0000-0002-2113-4863]{Weiguang Cui}
\affiliation{Universidad Autónoma de Madrid: Madrid, Madrid, ES}
\affiliation{Centro de Investigación Avanzada en Física Fundamental (CIAFF), Universidad Autónoma de Madrid, Madrid~28049, Spain}
\affiliation{Institute for Astronomy, University of Edinburgh, Blackford Hill, EH9 3HJ, Edinburgh, UK}
\email{} 

\author[orcid=0000-0001-8919-7409]{Yun Chen}
\affiliation{National Astronomical Observatories, Chinese Academy of Sciences, Beijing 100101, China}
\affiliation{College of Astronomy and Space Sciences, University of Chinese Academy of Sciences, Beijing, 100049, China}
\email{} 

\author[orcid=0000-0003-2842-9434]{Romeel Dav\'e}
\affiliation{Institute for Astronomy, University of Edinburgh, Blackford Hill, EH9 3HJ, Edinburgh, UK}
\affiliation{Department of Physics and Astronomy, University of the Western Cape, Robert Sobukwe Rd, 7535, Cape Town, South Africa}
\email{} 

\begin{abstract}
Satellite galaxies in massive halos are seen to be preferentially quenched along the major axes of their central galaxies.  In contrast, the AGN jet tends to point along the minor axis, which imposes interesting constraints on the role of AGN feedback in satellite galaxy quenching. Here we use the SIMBA, TNG100, EAGLE, and feedback-free SIMBA cosmological simulations at $z=0$ to show that this satellite anisotropy primarily originates from spatial segregation between quenched and unquenched populations. Quenched satellites are more centrally concentrated and more strongly aligned with the central galaxy's major axis, whereas unquenched satellites occupy larger radii and are more nearly isotropic. The persistence of this behaviour without stellar or active galactic nucleus feedback suggests that  feedback is not important for satellite quenching. Beyond the halo boundary, the signal varies strongly among simulations and correlates with the excess number of galaxies along the major axis. This links large-scale anisotropic quenching to the direction-dependent distribution of groups, clusters and filamentary environments. Our results provide a unified geometrical interpretation of anisotropic satellite quenching across halo and extra-halo scales.
\end{abstract}


\section{Introduction}\label{sec:intro}
Satellite galaxies with dark matter halos as the undissolved remnants of hierarchical structure formation are influenced by the triaxial halo shape and may still retain some information about large-scale structure. Meanwhile, the host dark matter halo provides a uniquely dense environment for their quenching process. For the former, it has been verified both in observation \citep{2004MNRAS.348.1236S,2005ApJ...628L.101B, 2006MNRAS.369.1293Y,2006ApJ...650..550A, 2008MNRAS.390.1133B} and in simulation \citep{2005MNRAS.363..146L,2005ApJ...624..505Z, 2007MNRAS.378.1531K,2008MNRAS.390.1133B, 2020MNRAS.491.5330S}. That satellite galaxies/subhalos preferentially trace the major axis of their host halo. In our companion study \citep{2026arXiv260427845Z}, we found that the structural tracer most closely associated with satellite anisotropy changes with distance from the central galaxy. On small scales, the satellite distribution follows the morphology of the central stellar component; across the host halo, it is more closely associated with the triaxiality of the dark-matter halo; and beyond the halo, it increasingly traces the surrounding filamentary structure. This spatial segregation, as well as the dynamical origin for the preferential major-axis distribution -- satellite trajectories also spend more time in major-axis-aligned regions -- provides the basis for the present study.

On the latter, satellite quenching records both the present host environment and the previous orbital and accretion history of the galaxy \citep{2009MNRAS.394.1213W, 2010ApJ...721..193P, 2012ApJ...757....4P, 2012MNRAS.424..232W, 2013MNRAS.432..336W}. Therefore, it has also been connected to different processes besides the feedback scheme, e.g. ram-pressure, tidal stripping and impulsive galaxy encounters, starvation or strangulation due to the removal or heating of the extended gas reservoir \citep{2005nfcd.conf..220M, 2008MNRAS.383..593M,
2012MNRAS.423.1277D,
2015MNRAS.454.2039F, 2016MNRAS.463.3083O, 2015ApJ...808L..27W, 2018MNRAS.475.3654Z, 
2023MNRAS.526.3716B}. Furthermore, this can also be connected to a slightly larger anisotropic infall region via pre-processing \citep{1999MNRAS.307..463B, 2012MNRAS.424..232W, 2013MNRAS.435.2713V, 2018MNRAS.475.3654Z} and contamination from these splashback galaxies. 

These two phenomena, anisotropic satellite positions and environmentally driven satellite quenching, are potentially connected by anisotropic satellite galaxy quenching (ASGQ), which describes the phenomenon where the quenched fraction of satellite galaxies along the minor axis of the central galaxy is lower than that along its major axis \citep{2021Natur.594..187M}. After the identification with the Sloan Digital Sky Survey data by \citet{2021Natur.594..187M}, this observational signal has subsequently been detected over a wider range of halo masses, radii and cosmic epochs. Using CLASH clusters, \citet{2022MNRAS.511.2659S} found redder and more strongly quenched satellite populations near the major axes of brightest cluster galaxies out to $z\simeq0.5$. With the much larger HSC-SSP cluster sample, \citet{2023MNRAS.519...13A} detected ASGQ over $0.25<z<1$ and found the anisotropy to be clearer inside $R_{200\mathrm m}$ \footnote{$R_{200\mathrm m}$ denotes the radius enclosing a mean mass density of 200 times the cosmic mean density.} than in the cluster outskirts. More recently, CLASH measurements have traced the signal to approximately $3R_{200}$ \footnote{The median value of $R_{200}$ quoted here is about 933 kpc.}, with a reported maximum near $1.25R_{200}$ and an interpretation based on pre-processing and anisotropic large-scale structure \citep{2025MNRAS.537.1542S}. A new study of spectroscopically confirmed clusters at $0.9<z<1.4$ reports a modest-significance detection beyond $z=1$ and argues, using accretion histories and a delay-then-rapid quenching model, that direction-dependent pre-processing can reproduce the signal \citep{2026arXiv260623790A}.

Despite these advances, the fundamental physical origin of the ASGQ effect remains an open and debated question. Besides the explanation in \citet{2021Natur.594..187M}, which suggests the AGN feedback as the cause, several work proposed different answers: By applying UniverseMachine to the SMDPL simulation, \citet{2023ApJ...949L..13K} demonstrated that ASGQ can arise in the absence of anisotropic baryonic feedback. In that model, satellites near the major axis tend to have earlier accretion times and larger peak subhalo masses, making anisotropic quenching a natural consequence of hierarchical assembly. Analysis of IllustrisTNG similarly indicates that young satellites preferentially enter along the central major axis, which tends to align with the feeding filament, whereas older satellites have been dynamically mixed and no longer retain an equally strong ASGQ signature \citep{2025A&A...693A.113Z}. Observational work on massive clusters has instead emphasised anisotropic pre-processing, finding that galaxy density decreases less rapidly along the BCG major axis and that the passive fraction remains higher in that direction even at fixed local projected density \citep{2025MNRAS.537.1542S}. These existing suggestions for the formation of ASGQ: direction-dependent modification of circumgalactic gas by central AGN feedback; anisotropic accretion histories coupled to satellite quenching; and spatially anisotropic pre-processing in filaments, groups and neighbouring clusters, exclude a simply possibility: an ASGQ signal can arise when quenched and unquenched satellites have different radial distributions. Moreover, the connection between the intrinsic satellite anisotropy distribution and ASGQ signal is unclear. As shown in our companion paper  \citep{2026arXiv260427845Z}, the dominant structural tracer of satellite anisotropy changes from the central galaxy to the host halo and then to the filamentary environment, the physical origin of ASGQ need not be the same inside and outside the halo.

Here we test these possibilities using four hydrodynamical cosmological simulation suites: SIMBA, TNG100, EAGLE and a feedback-free SIMBA calculation. We measure ASGQ as a joint function of satellite stellar mass and central-galaxy distance, covering both intra-halo and extra-halo scales. Within $R_{200\mathrm c}$ \footnote{$R_{200\mathrm c}$ is the radius enclosing a mean mass density of 200 times the cosmic critical density.}, we show that ASGQ is associated with spatial segregation between quenched and unquenched satellites: quenched satellites are more centrally concentrated, and their distribution consequently trace the major-axis of the central galaxy more strongly, whereas the spatial distribution of unquenched satellites are more radially extended and more nearly isotropic. The persistence of this behaviour in the feedback-free run demonstrates that stellar and AGN feedback are not required to generate the intra-halo signal. Outside the halo, we show that ASGQ varies substantially among simulations and correlates with the directional galaxy-number asymmetry. This links extra-halo ASGQ to the anisotropic distribution of groups, clusters and filamentary environments and provides a unified, scale-dependent framework for interpreting the phenomenon.

\section{Methods} \label{sec:methods}
\subsection{Simulation Data and Sample Selection} \label{sec:data}

Our primary analysis utilizes simulation data at $z = 0.0$ from the SIMBA-m100n1024 suite \citep{2019MNRAS.486.2827D}, complemented by high-resolution snapshots from TNG100-1 \citep{2019ComAC...6....2N},  EAGLE-L100N1504 \citep{2015MNRAS.446..521S} and SIMBA-m50n512 nofb simulation. While SIMBA, TNG100, and EAGLE incorporate comprehensive sub-grid baryonic physics, they differ significantly in their AGN feedback and black hole accretion implementations \citep{Vogelsberger2020, Dave2020, Habouzit2021}:
\begin{itemize}
    \item \textbf{SIMBA} implements torque-limited/Bondi accretion with kinetic outflows in both quasar and jet modes, generating bipolar AGN outflows with a preferred feedback axis \citep{2019MNRAS.486.2827D}.\\
    \item \textbf{TNG100} combines Bondi accretion with hybrid thermal-kinetic feedback, where the kinetic feedback is randomised in direction for each injection event \citep{2017MNRAS.465.3291W}.\\
    \item \textbf{EAGLE} adopts a Bondi-based accretion model with isotropic thermal feedback \citep{2015MNRAS.446..521S}.
\end{itemize}

These distinct implementations provide diverse AGN feedback prescriptions, allowing us to assess the robustness of our results across different simulation models. To explicitly isolate the impact of feedback mechanisms, we also analyse the SIMBA-nofb run \citep{2024MNRAS.527.1612Y}, which intentionally disables stellar feedback, AGN feedback, and X-ray heating.
Halos and galaxies are identified using the 6D Friends-of-Friends algorithm via \texttt{CAESAR}\footnote{\tt https://caesar.readthedocs.io/en/latest/} package, followed by the implementation of data screening based on these outputs. Relevant information on the simulation datasets employed herein is summarized in Table \ref{tab:params_simulation}. 

\begin{table}[h!]
    \hspace*{-1.5cm}
    \centering
    \setlength{\tabcolsep}{5pt}  
    \renewcommand{\arraystretch}{1.5}  
    \begin{tabular}{lcccccccc}
        \toprule
        Full name (Abbreviation in this paper) & & $L_\mathrm{box} \, [\mathrm{Mpc}]$ & $N_\mathrm{DM}$ & & $m_\mathrm{gas} \, [M_{\odot}]$ & & $m_\mathrm{DM} \, [M_{\odot}]$\\
        \midrule
        SIMBA-m100n1024 (SIMBA) & & 147.1 & $1024^3$ & & $1.82\times10^{7}$ & & $9.60\times10^{7}$\\
        TNG100-1 (TNG100) & & 110.7 & $1820^3$ & & $1.40\times10^{6}$ & & $7.50\times10^{6}$\\
        EAGLE-L100N1504 (EAGLE) & & 100.0 & $1504^3$ & & $1.81\times10^{6}$ & & $9.70\times10^{6}$\\
        SIMBA-m50n512 nofb  (SIMBA-nofb) & & 73.5 & $512^3$ & & $1.82\times10^{7}$ & & $9.60\times10^{7}$\\
        \bottomrule
    \end{tabular}
    \caption{Physical parameters for three fiducial simulations and one feedback-deficient simulation. From left-to-right the columns show: simulation name suffix; comoving box length; number of dark matter particles; initial gas particle mass; dark matter particle mass.}
    \label{tab:params_simulation}
\end{table}

\begin{figure}[h!]

\centering
\includegraphics[width=0.7\textwidth]{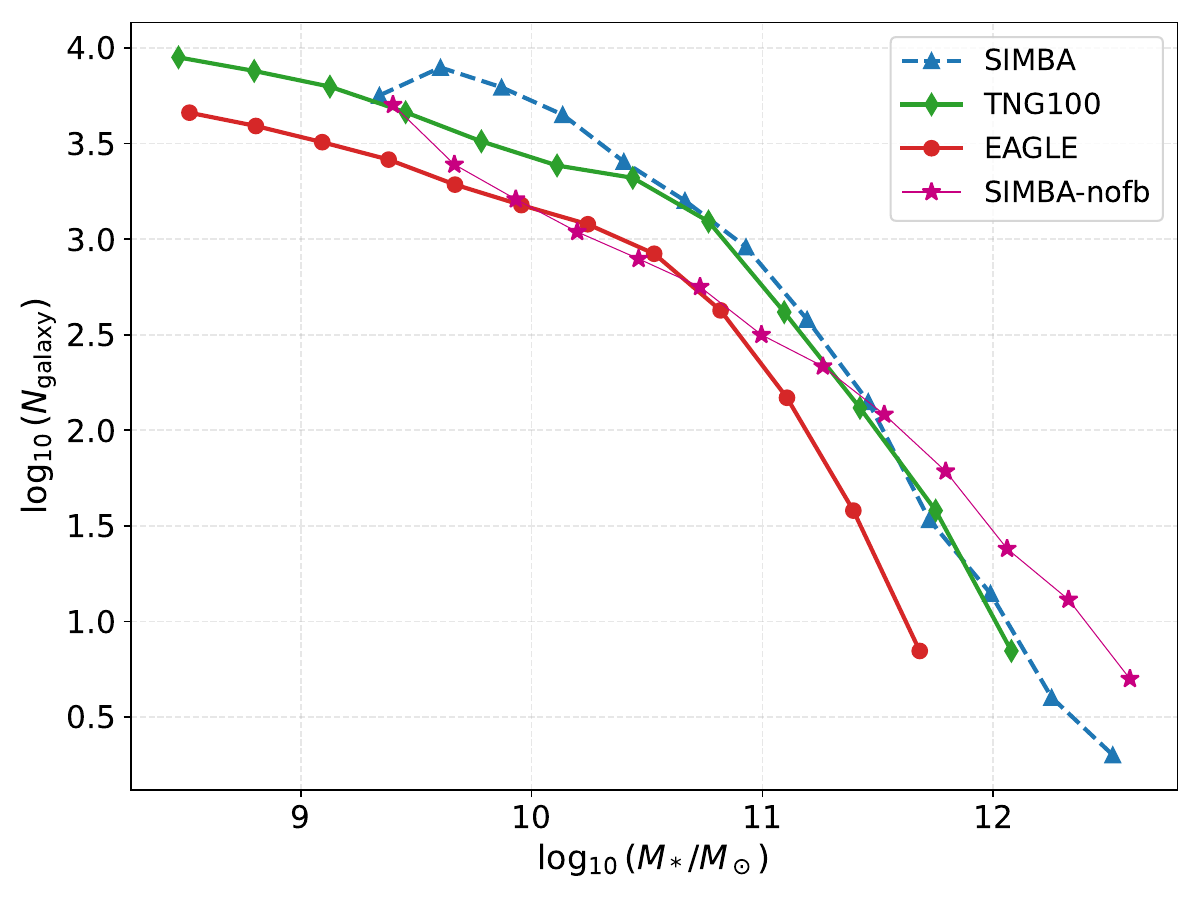}

\caption{Stellar mass distributions of galaxies at $z=0$ from the SIMBA, TNG100, EAGLE, and SIMBA-nofb simulations. Low-mass galaxies dominate the full sample across all runs, in line with the bottom-up hierarchical scenario of cosmic structure assembly.}
\label{fig:starmass_function}
\end{figure}

For the sample selection, we impose a minimum halo mass threshold of $M_{200\rm c} = 10^{12}\,\mathrm{M}_\odot$ with central galaxies containing at least 100 stellar particles. Additionally, satellite galaxies are required to have stellar masses above $100 \times m_\mathrm{gas}$ and spatial distances greater than twice the half-mass radius of the central galaxy ($2r_{\mathrm{c,\,half}}$) to ensure numerical reliability. It is important to clarify that the satellite galaxies considered in this work are not required to be gravitationally bound, and thus have a broader connotation than satellite galaxies in the general sense. Figure \ref{fig:starmass_function} presents the stellar mass distribution of the galaxy in different simulations. We find that the shapes of the curves for the three fiducial simulations are in good agreement, whereas the no-feedback run produces an excess of massive galaxies. This likely reflects the important role of the baryonic feedback processes in suppressing these massive structures. Further note that the normalisation difference between the three simulations is not only due to the simulation box size, but also to their intrinsic baryon models.

\subsection{Quenched Galaxy Classification}

To separate unquenched and quenched galaxies, , we follow the criterion established by \citet{2024MNRAS.527.1612Y}, which is based on the observed star-forming main sequence (SFMS) reported by \citet{2018MNRAS.477.3014B}. Specifically, the boundary dividing the two populations is defined as:

\begin{equation}
\log(\mathrm{SFR} / \mathrm{M}_\odot\,\mathrm{yr}^{-1}) = 0.73 \log(M_* / \mathrm{M}_\odot) - 8.33
\end{equation}

\noindent where $\mathrm{SFR}$ denotes the star formation rate and $M_*$ represents the stellar mass of the galaxy. Galaxies lying below this main-sequence relation are classified as quenched, whereas those residing on or above it are identified as unquenched.

\begin{figure}[h!]

\centering
\includegraphics[width=0.7\textwidth]{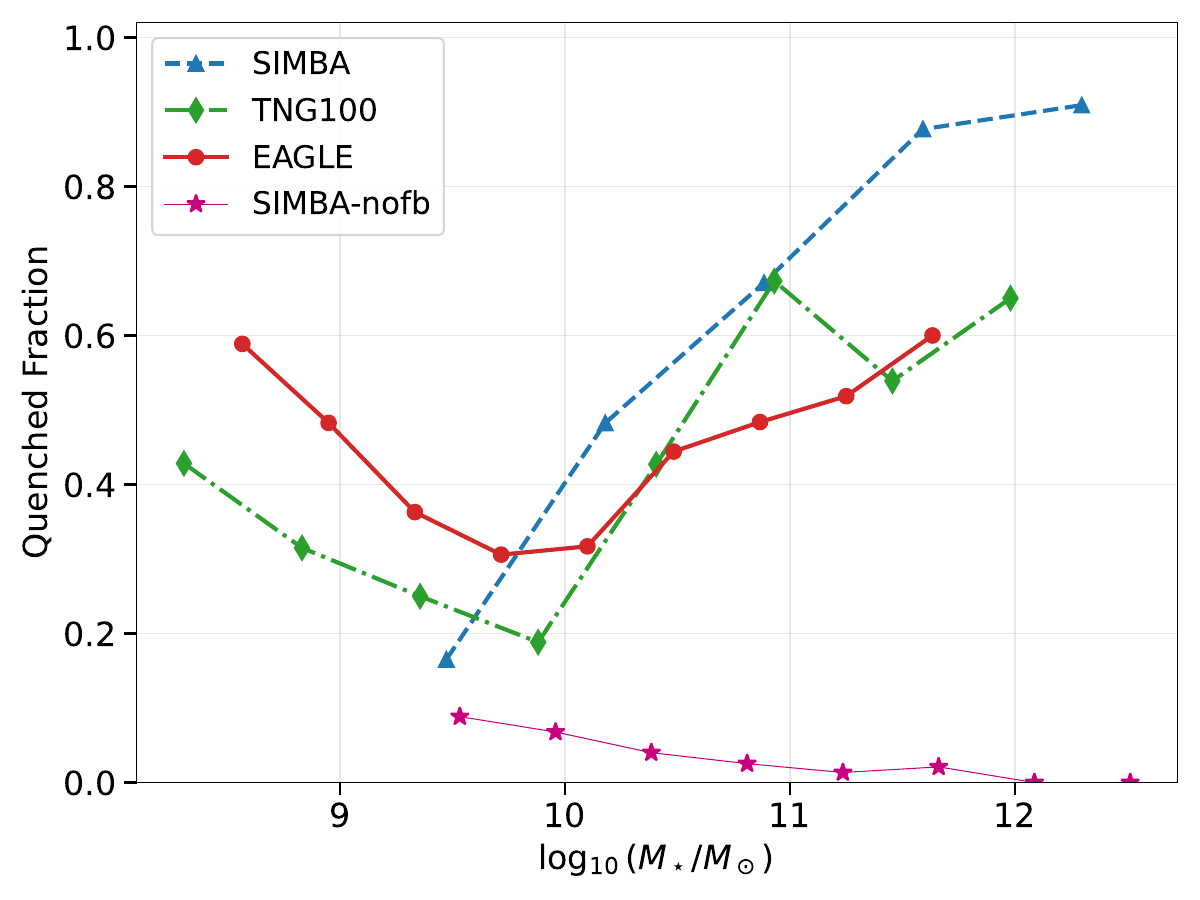}

\caption{Quenched fraction of satellite galaxies versus stellar mass for the SIMBA, TNG100, EAGLE, and SIMBA-nofb simulations. The three full-physics runs (SIMBA,  TNG100, EAGLE) yield similar trends in the stellar-mass dependence of the quenched fraction.}
\label{fig:F_Q}
\end{figure}

Based on this classification, we calculate the quenched fraction as a function of stellar mass across four simulation suites: SIMBA, TNG100, EAGLE, and SIMBA-nofb. As illustrated in Figure \ref{fig:F_Q}, SIMBA, TNG100, and EAGLE simulations exhibit consistent qualitative trends in their quenched fractions with respect to $M_*$. In contrast, the SIMBA-nofb run shows a significantly suppressed quenched fraction for massive galaxies, remaining at very low levels due to the absence of effective feedback mechanisms to quench star formation. 

\subsection{Orientation Measurement and Satellite Statistical Counting}
\label{subsec:count}

We follow the orientation-measurement and directional-counting procedure
introduced in our companion study \citep{2026arXiv260427845Z}. The principal
axes of each central galaxy are obtained from a randomized principal-component analysis
of its stellar-particle positions. We then construct symmetric bicones centred
on the central galaxy with a half-opening angle of $\theta=45^{\circ}$.
Separate bicones are aligned with the major and minor principal axes \footnote{Throughout this work, ``major axis'' and ``minor axis'' refer to the principal
axes of the central stellar distribution. They do not refer directly to the
principal axes of the dark-matter halo or to the orientation of a cosmic
filament.}.

Quenched, unquenched and total galaxy counts are measured within each
directional selection. Counts are first obtained for individual host systems
and are then summed over the selected host sample when constructing stacked
measurements. Equations~\ref{eq:FQ_definition} and
\ref{eq:DeltaFQ_definition} define the directional quenched fractions and the
ASGQ amplitude.

\subsection{Uncertainty estimation}
\label{subsec:uncertainties}

For each central galaxy, we obtain 50 independent sets of triaxial ellipsoid axes using the method described in Section \ref{subsec:count}. Stacking central galaxies accordingly yields 50 distinct stacked realizations, each producing one curve. These 50 stacked curves together form a scatter band. The error bars shown in previous figures correspond to the interpolated sample quantiles spanning $0.15\%$ to $99.85\%$ (approximately corresponding to a $3\sigma$ confidence interval). In Section \ref{sec:extra_origins}, by contrast, we only present results computed from the averaged triaxial‑ellipsoid axis orientations of each central galaxy, obtained by averaging over the 50 independent sets.

\section{Anisotropic quenching in simulations}

\begin{figure}[h!]
\includegraphics[width=0.5\textwidth]{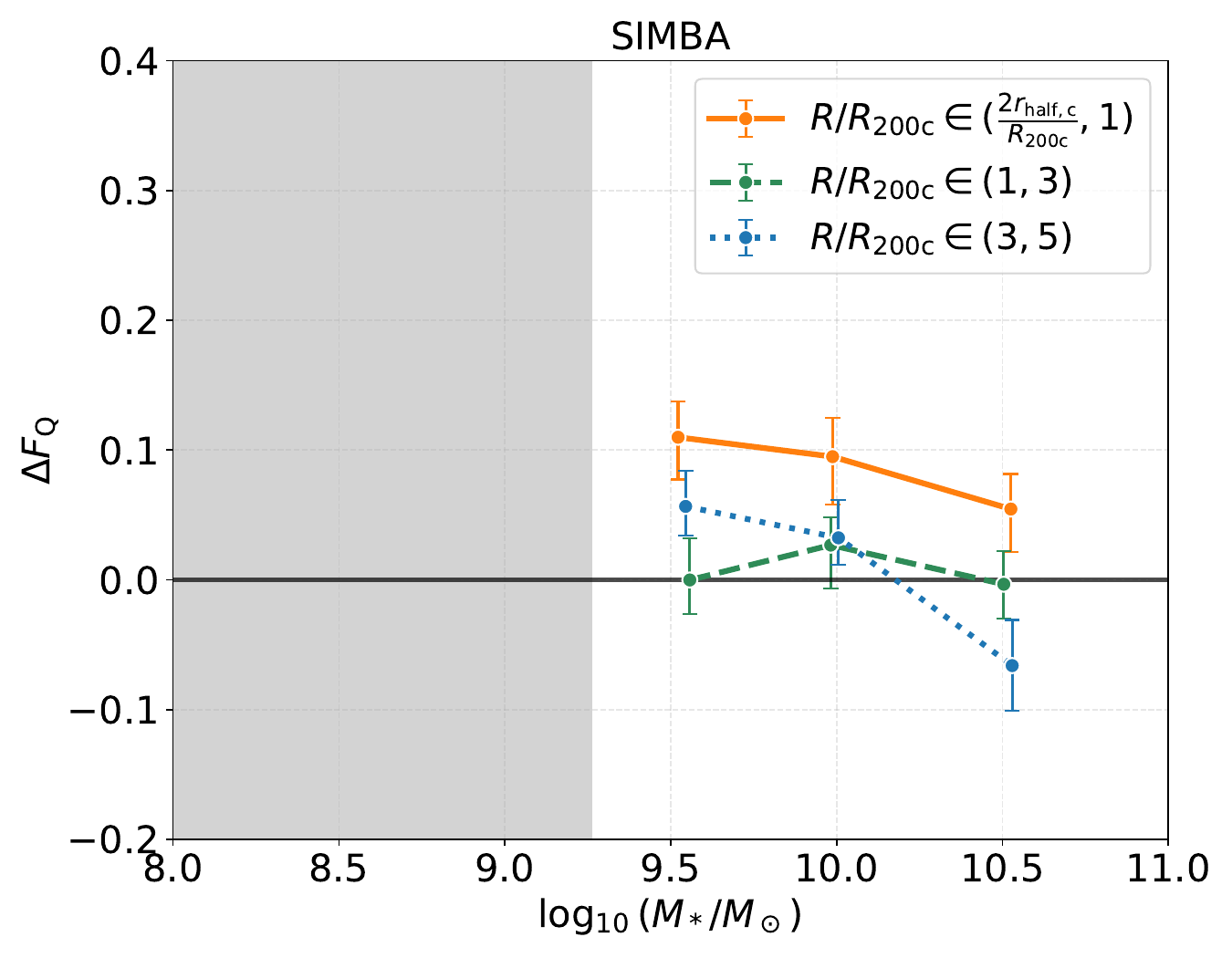}
\hspace{5pt}
\includegraphics[width=0.5\textwidth]{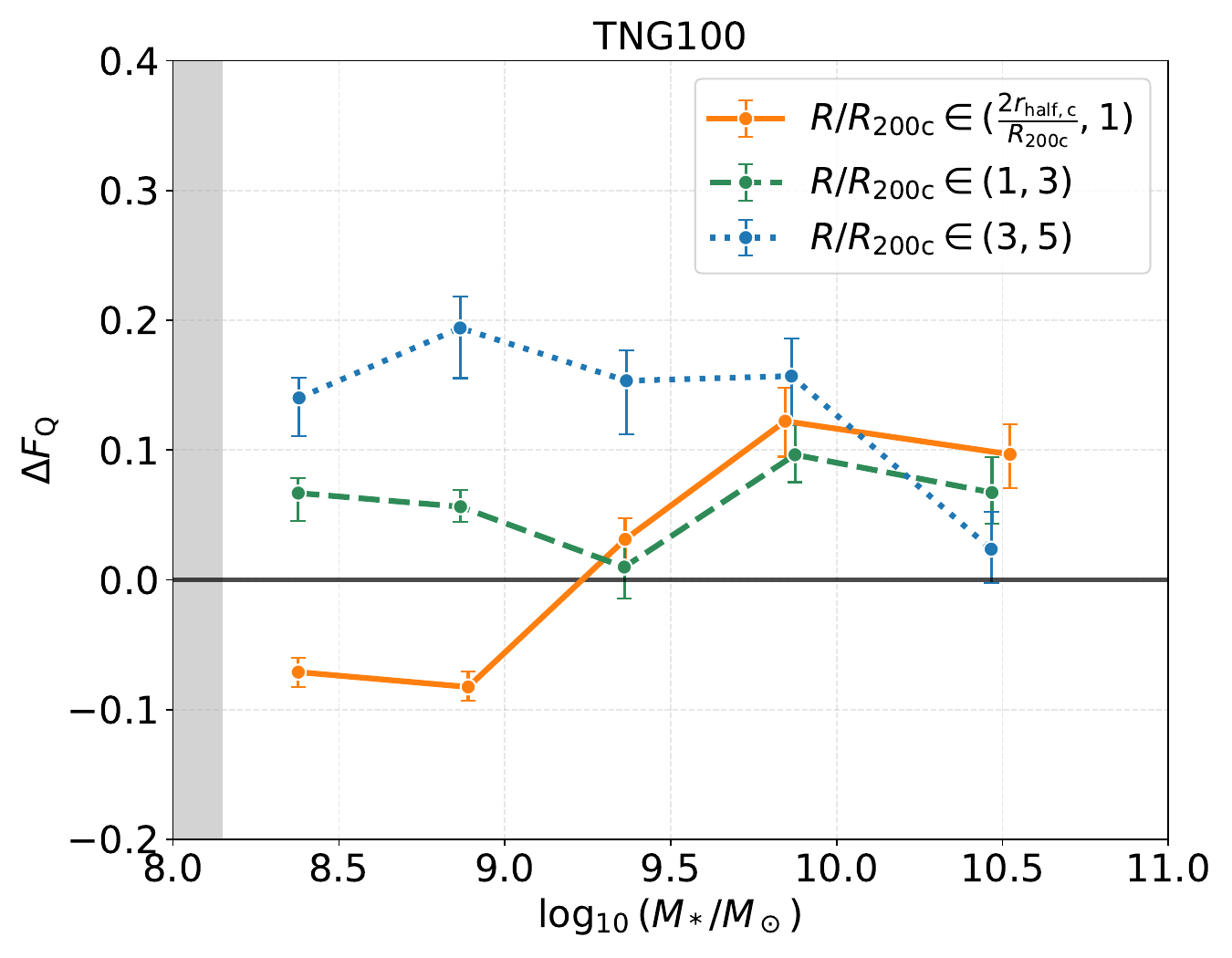}
\vspace{7pt}
\includegraphics[width=0.5\textwidth]{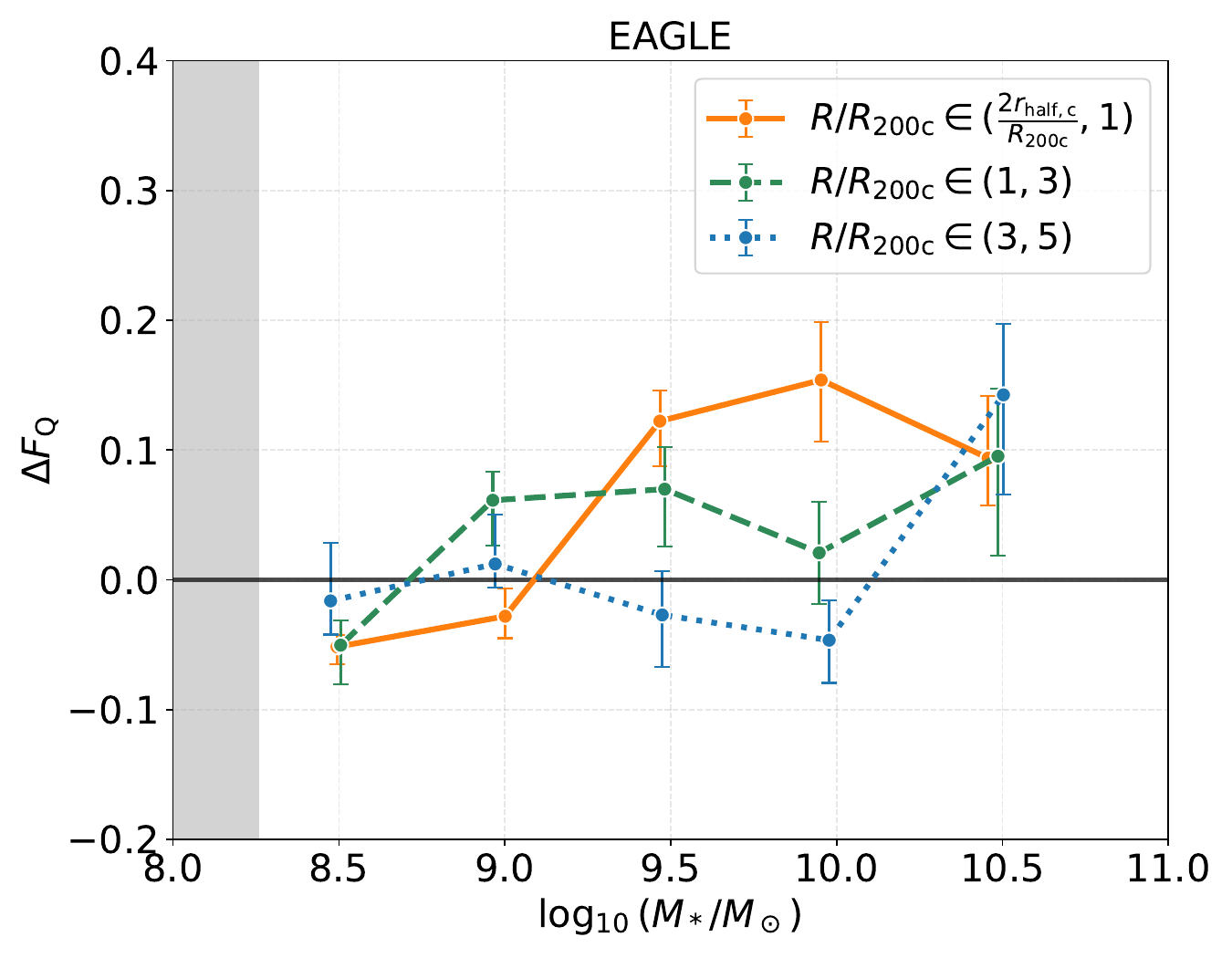}
\hspace{5pt}
\includegraphics[width=0.5\textwidth]{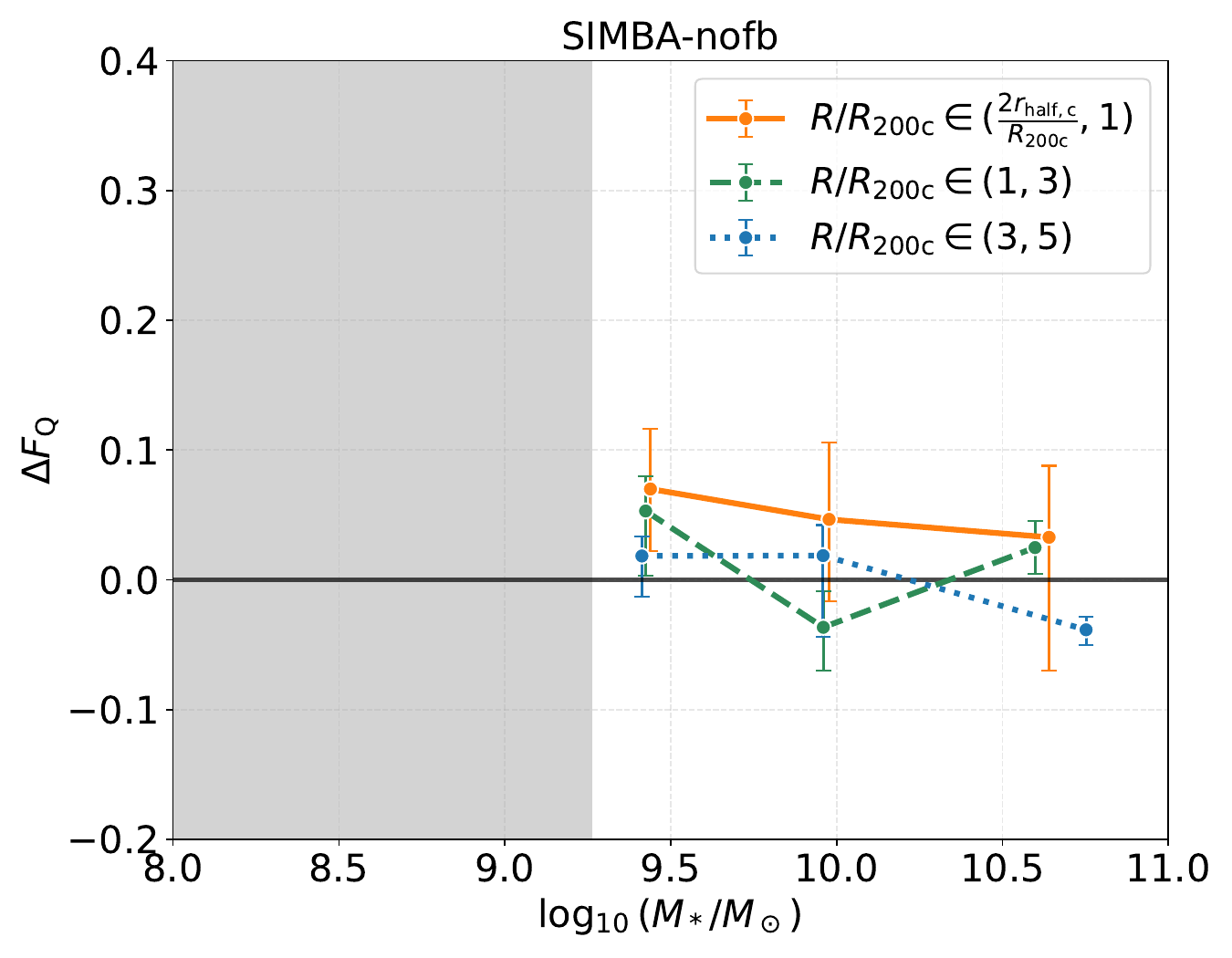}
\caption{
\textbf{Stellar-mass and radial dependence for the ASGQ signal.} The quenched-fraction difference, $\Delta F_{\mathrm{Q}} =F_{\mathrm{Q,major}}-F_{\mathrm{Q,minor}}$, is shown as a function of satellite stellar mass for SIMBA, TNG100, EAGLE, and SIMBA-nofb simulations. Different curves correspond to the radial intervals indicated in each panel.  The error bands represent the $3\sigma$ confidence level (encompassing 99.7\% of the statistical sample). The grey region marks stellar masses below the resolution threshold, $M_{\star}<100\,m_{\mathrm{gas}}$. Within $R_{200\mathrm{c}}$, satellites with $M_{\star}\geq10^{9.5}\,\mathrm{M}_{\odot}$ consistently produce a positive ASGQ signal ($\sim 0.1$) in all four simulations.}
\label{fig:ASGQ_Mstar_R}
\end{figure}

\begin{figure}[h!]
\centering
\includegraphics[width=0.47\textwidth]{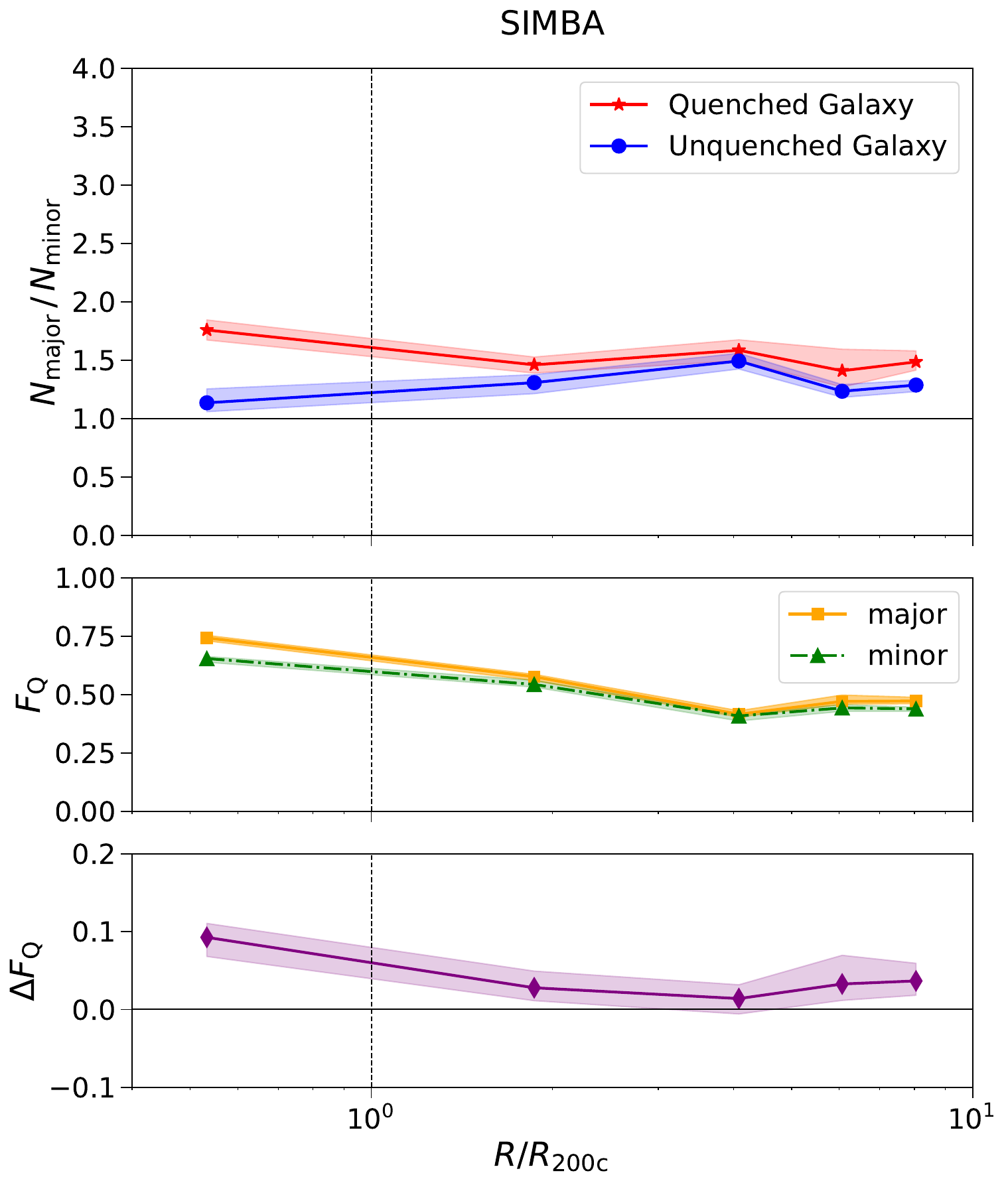}
\hspace{0pt}
\includegraphics[width=0.47\textwidth]{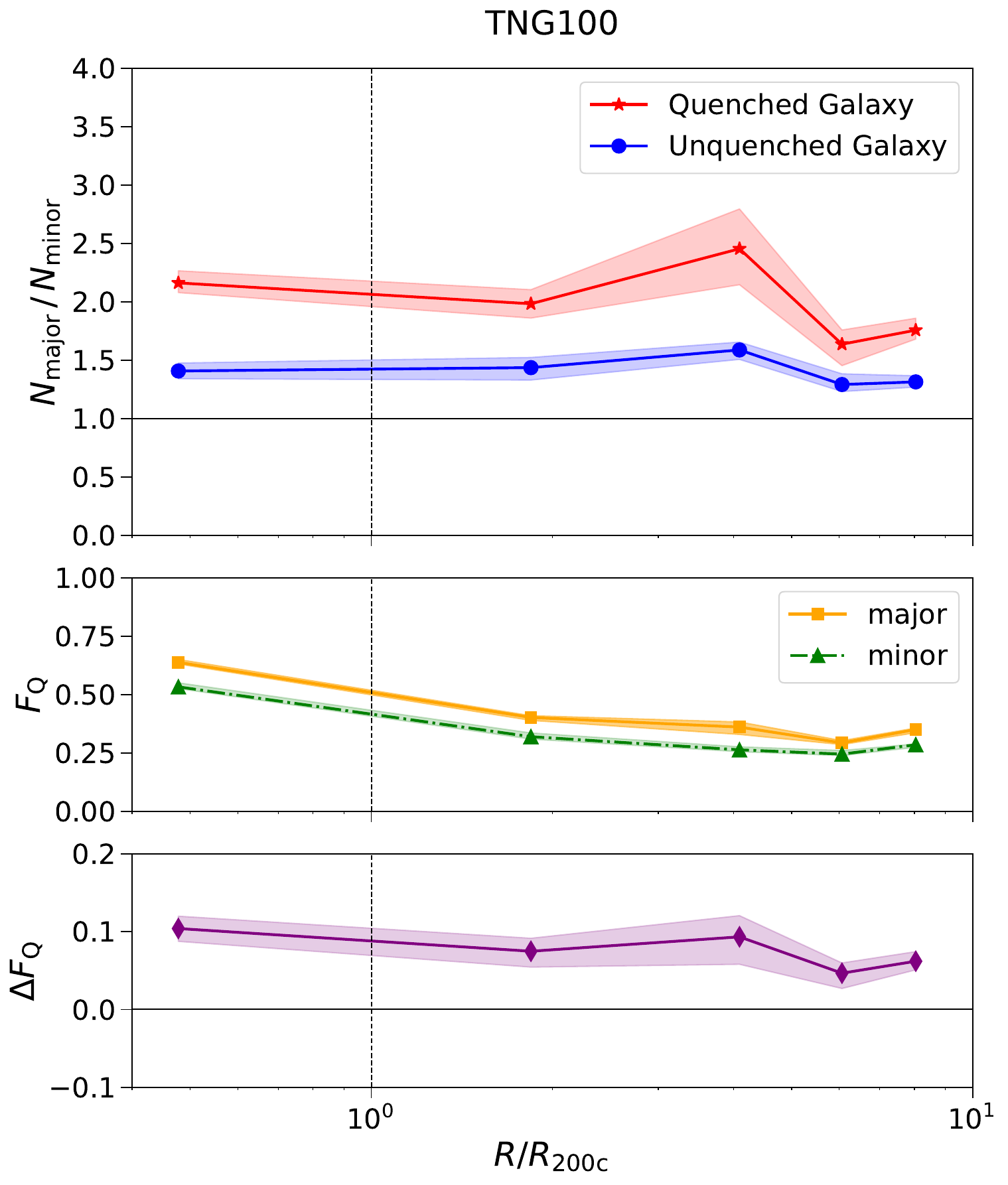}
\vspace{10pt}
\includegraphics[width=0.47\textwidth]{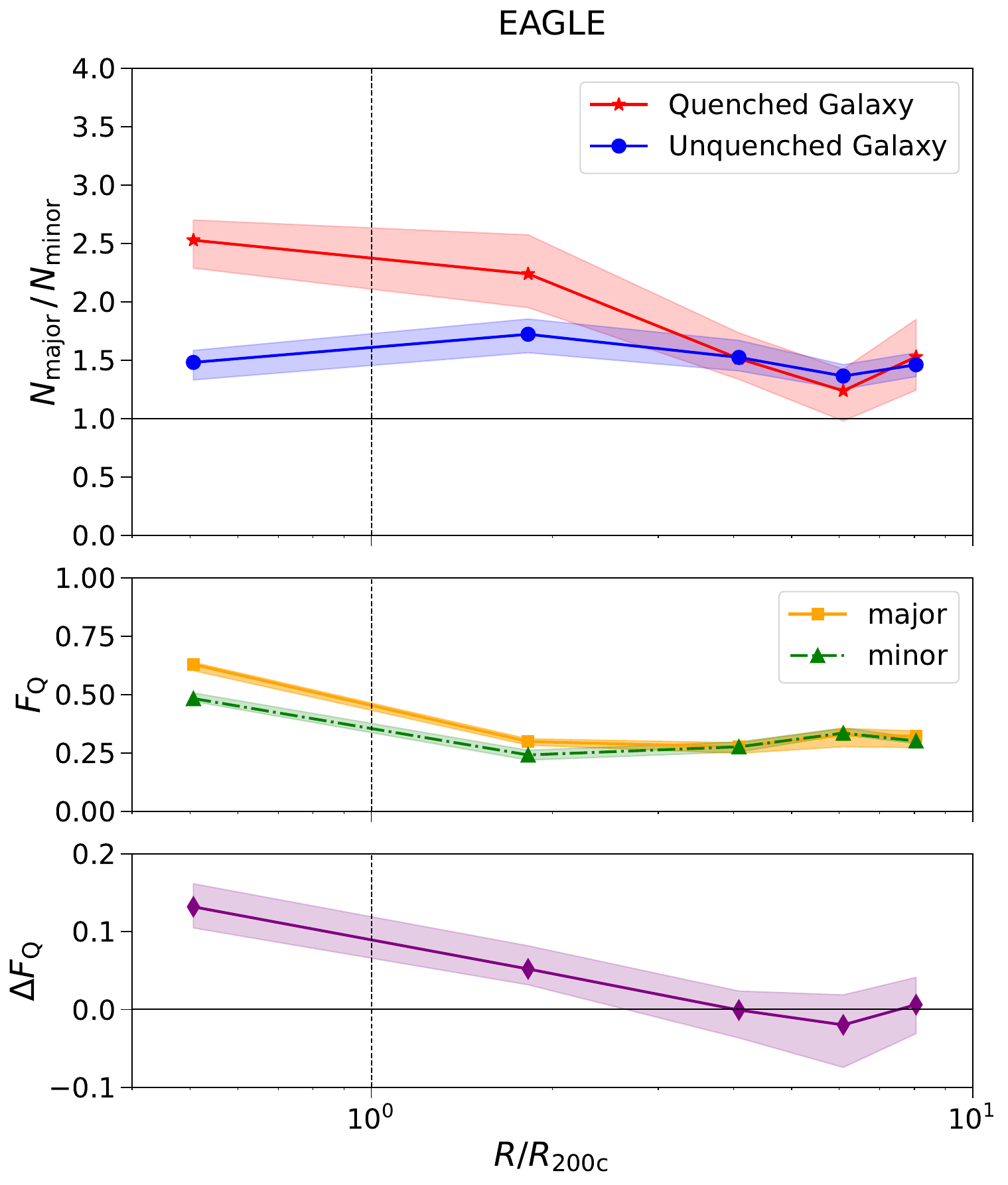}
\hspace{0pt}
\includegraphics[width=0.47\textwidth]{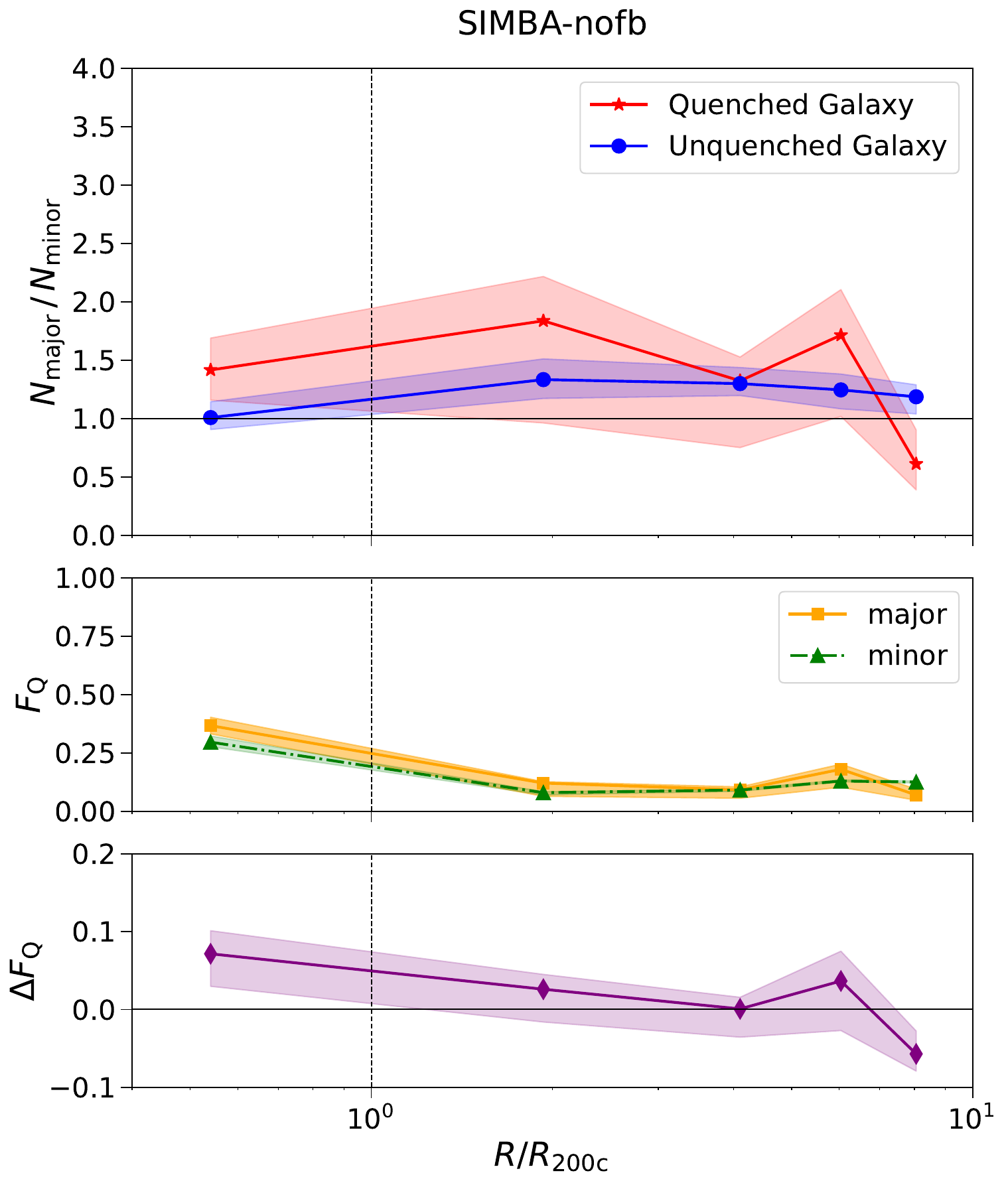}
\caption{
\textbf{Radial dependence of the satellite galaxy statistics and ASGQ signal.} Results are shown for satellites with $M_{\star}\geq10^{9.5}\,\mathrm{M}_{\odot}$ in SIMBA, TNG100, EAGLE and SIMBA-nofb. Top: major-to-minor axis number ratios for quenched and unquenched satellites. Middle, radial quenched fractions profile. Bottom, the ASGQ signal, $\Delta F_{\mathrm{Q}}$ as function of $R_{200c}$. Horizontal reference lines mark an isotropic number ratio in the top row and a zero quenched-fraction difference in the bottom row. Shaded error bands denote the $3\sigma$ confidence interval. The measured ASGQ signal is intrinsically tied to the differing spatial distributions of quenched versus unquenched satellite populations.}
\label{fig:QF_R_BCG}
\end{figure}

We follow the observation results by using the central galaxies' axes to identify the satellite galaxies lying in the major versus minor axes. The central galaxies' major and minor axes are identified based on a PCA-based orientation measurement, which is presented in \citet{2026arXiv260427845Z} to quantify the spatial anisotropy of satellite galaxies. Specifically, we only use the star particles' positions of the central galaxy to measure its triaxial ellipsoid orientations. Then, a symmetric bicone with an angle of $\theta=45^\circ$ centred at the central galaxy is used to ensure sufficient satellite sample statistics. We tally satellite numbers within each axis-aligned bicone for individual systems and aggregate the counts across the full sample. Additionally, we consider the uncertainty in determining galaxy orientation. Because satellite counts depend on the adopted orientation, our statistical uncertainties account solely for errors in the measured principal axes of the central galaxy. The complete technical details of the entire pipeline are provided in the aforementioned work.

For a given direction $\boldsymbol{\vec n}$, the quenched fraction is defined as \begin{equation} 
F_{\mathrm{Q}}(\boldsymbol{\vec n}) = \frac{N_{\mathrm{Q}}(\boldsymbol{\vec n})} {N_{\mathrm{tot}}(\boldsymbol{\vec n})}, \qquad \boldsymbol{\vec n} \in \{\mathrm{major},\mathrm{minor}\}, \label{eq:FQ_definition} 
\end{equation} 
where $N_{\mathrm{Q}}(\boldsymbol{\vec n})$ and $N_{\mathrm{tot}}(\boldsymbol{\vec n})$ are the quenched and total galaxy counts, respectively, within the corresponding directional bicone. We characterise the ASGQ signal using 
\begin{equation} 
\Delta F_{\mathrm{Q}} = F_{\mathrm{Q,major}} - F_{\mathrm{Q,minor}}. \label{eq:DeltaFQ_definition} 
\end{equation}
A positive value of $\Delta F_{\mathrm{Q}}$ therefore indicates that the quenched fraction is higher along the major axis of the central galaxy than along its minor axis. 

Figure~\ref{fig:ASGQ_Mstar_R} shows $\Delta F_{\mathrm{Q}}$ as a function of satellite stellar mass in three radial intervals: $\frac{2r_\mathrm{half, c}}{R_{200\mathrm{c}}}<R/R_{200\mathrm{c}}<1$, $1<R/R_{200\mathrm{c}}<3$ and $3<R/R_{200\mathrm{c}}<5$, where $r_\mathrm{half, c}$ is the half total mass radius of the central galaxies. The detailed mass and radial dependences differ among these simulations, particularly outside the halo and at low satellite stellar mass. This variation indicates that the ASGQ signal is sensitive to the simulation features and is not universal across all mass and radial ranges. Nevertheless, the four simulations share one notable result: within $R_{200\mathrm{c}}$, satellites with $M_{\star}\geq10^{9.5}\,\mathrm{M}_{\odot}$ consistently exhibit a positive ASGQ signal with slightly different stellar mass dependence, a characteristic amplitude of approximately 0.1. This stellar-mass range lies above the numerical-resolution threshold in all four simulations, which forms the main sample for investigating the physical origin of the ASGQ signal. However, extra-halo results are less consistent: In TNG100, the ASGQ signal of low-mass galaxies increases towards larger central-galaxy distances, reaching $\sim M_* = 10^ {10} M_\odot$ and reversing at the highest mass bin, in qualitative agreement with the extended signal previously reported for this simulation \citep{2025A&A...693A.113Z}. SIMBA shows a closer evolution along the satellite galaxy stellar mass, but the ASGQ signal is much weaker compared to TNG100. EAGLE also has a weaker ASGQ signal, but the mass dependence trend is reversed or shifted. For SIMBA-nofb, the ASGQ signal is almost consistent with zero, which corresponds to the absence of stellar feedback, active galactic nucleus (AGN) feedback and X-ray heating in its run. These differences could reflect the combined effects of baryonic modelling, simulation volume, satellite demographics, and numerical resolution. We therefore restrict our primary physical interpretation of the intra-halo signal to satellites with $M_{\star}\geq10^{9.5}\,\mathrm{M}_{\odot}$ and provide the low-mass samples separately within the following Methods section (Section~\ref{sec:methods}) as a methodological consistency test. Importantly, a positive intra-halo signal is also present in SIMBA-nofb, despite the absence of stellar feedback, active galactic nucleus feedback, and X-ray heating in its run. This comparison demonstrates that these feedback channels are not necessary for generating the measured intra-halo ASGQ signal. The result is consistent with previous work showing that ASGQ can arise without anisotropic feedback through direction-dependent satellite assembly histories \citep{2023ApJ...949L..13K}.

More detailed information on the satellite galaxy distributions and properties are presented in Figure~\ref{fig:QF_R_BCG} for all galaxies with  $M_{\star}\geq10^{9.5}\,\mathrm{M}_{\odot}$. The top row shows the major-to-minor number ratios of quenched and unquenched satellites. Within $R_{200\mathrm{c}}$, the quenched population has a systematically larger directional ratio than the unquenched population in all four simulations. The difference decreases with radius and becomes less consistent between these simulations outside the halo. While the unquenched galaxies seems to be radial independent for all simulations. The middle row shows the quenched fractions measured separately along the major and minor axes. For all simulations, these quenched fractions in both major and minor axes are highest near the central galaxy and decreases with radial distance. Within the halo, $F_{\mathrm{Q,major}}$ exceeds $F_{\mathrm{Q,minor}}$ for all simulations. However, the two profiles approach each other at larger radii following different trends among these simulations. This implies that the angular dependence of satellite quenching weakens or vanishes, with the outcome dependent on both radius and simulation. The bottom row shows the resulting $\Delta F_{\mathrm{Q}}$ profiles, which are simply the difference between the two profiles in the middle panel. This signal shows the strongest ASGQ signal closer to the central galaxy with the highest amplitude of approximately 0.1 for all simulations, and decreases towards larger radii at different pace. 

Taken together, Figure~\ref{fig:QF_R_BCG} shows that the presence of the ASGQ signal and a higher ratio of major-axis to minor-axis galaxy counts for quenched satellites than for unquenched satellites occur together, indicating a significant correlation between the ASGQ signal and major-axis anisotropy. As the ASGQ effect can be succinctly defined via the inequality
\begin{equation}
F_{\rm Q,{\rm major}} > F_{\rm Q,{\rm minor}}, \label{eq:FQ_inequality}
\end{equation}
which is equivalent to
\begin{equation} 
\frac{N_{\mathrm{Q,major}}}{N_{\mathrm{major}}} > \frac{N_{\mathrm{Q,minor}}}{N_{\mathrm{minor}}}. \label{eq:anisotropy_radial} 
\end{equation}
Because the total directional counts as 
\begin{equation} 
N_{\mathrm{major}} = N_{\mathrm{Q,major}}+N_{\mathrm{U,major}}, \qquad N_{\mathrm{minor}} = N_{\mathrm{Q,minor}}+N_{\mathrm{U,minor}}, \label{eq:N_decomposition} 
\end{equation} 
where the subscripts Q and U denote quenched and unquenched galaxies, respectively, the condition for a positive ASGQ signal becomes 
\begin{equation} 
\frac{N_{\mathrm{Q,major}}} {N_{\mathrm{Q,major}}+N_{\mathrm{U,major}}} > \frac{N_{\mathrm{Q,minor}}} {N_{\mathrm{Q,minor}}+N_{\mathrm{U,minor}}}. \label{eq:FQ_inequality} 
\end{equation} 
For positive, non-zero directional counts, this condition is equivalent to \begin{equation} 
\frac{N_{\mathrm{Q,major}}}{N_{\mathrm{Q,minor}}} > \frac{N_{\mathrm{U,major}}}{N_{\mathrm{U,minor}}}. \label{eq:anisotropy_condition} 
\end{equation} 
Thus, ASGQ is present when the major-to-minor directional count ratio is larger for quenched satellites than for unquenched satellites. This relation is an algebraic consequence of the count definitions. This identifies the physical origin of the ASGQ signal, which must be established from the spatial distributions of the two populations.
However, it does not yet explain why the quenched and unquenched populations have different spatial anisotropies, which we will address in the next section by comparing their radial distributions.

\section{Spatial segregation of satellite quenching inside halos}

\subsection{Origin of the anisotropic satellite distribution}

\begin{figure}[h!]
\centering
\includegraphics[width=0.47\textwidth]{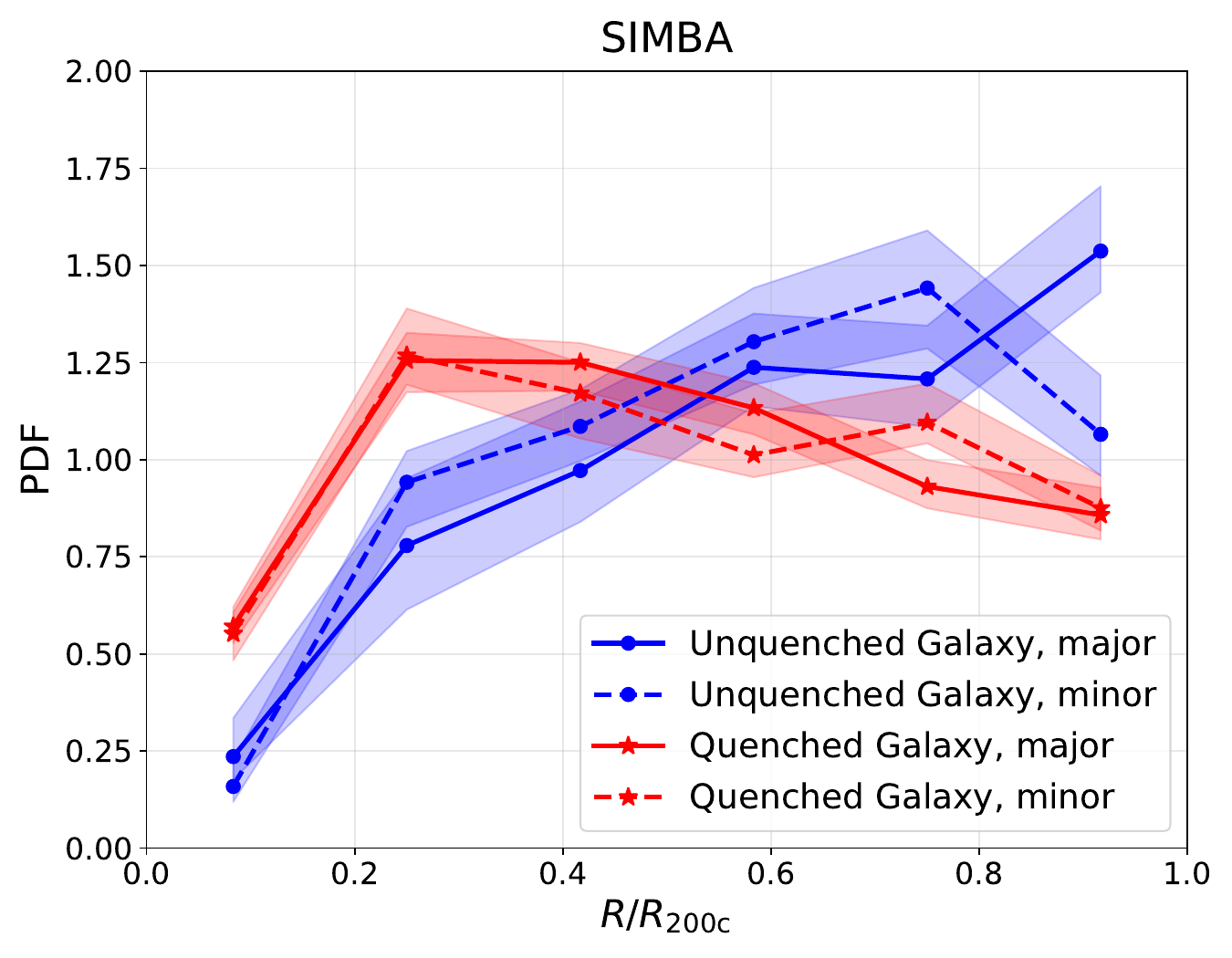}
\hspace{0pt}
\includegraphics[width=0.47\textwidth]{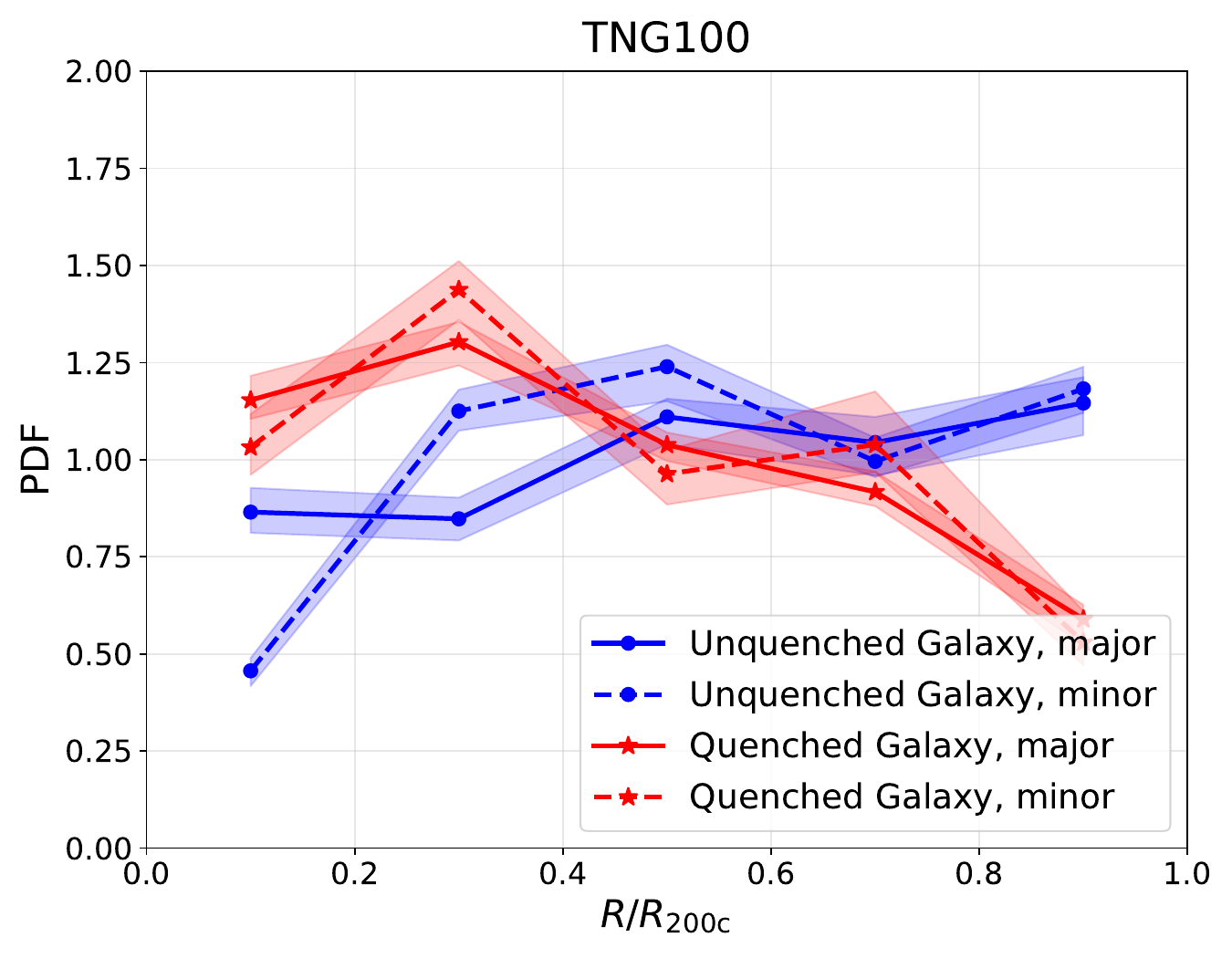}
\vspace{10pt}
\includegraphics[width=0.47\textwidth]{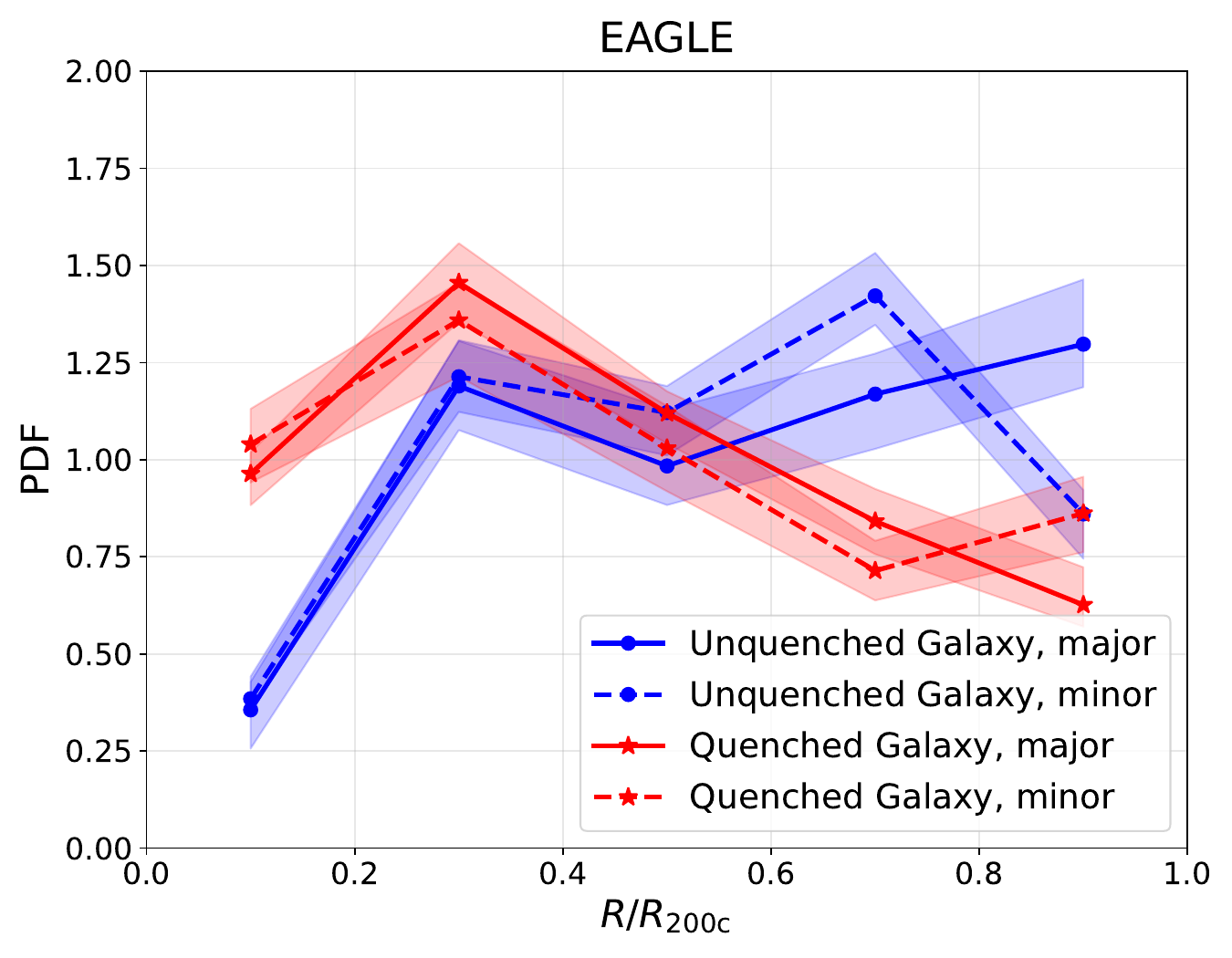}
\hspace{0pt}
\includegraphics[width=0.47\textwidth]{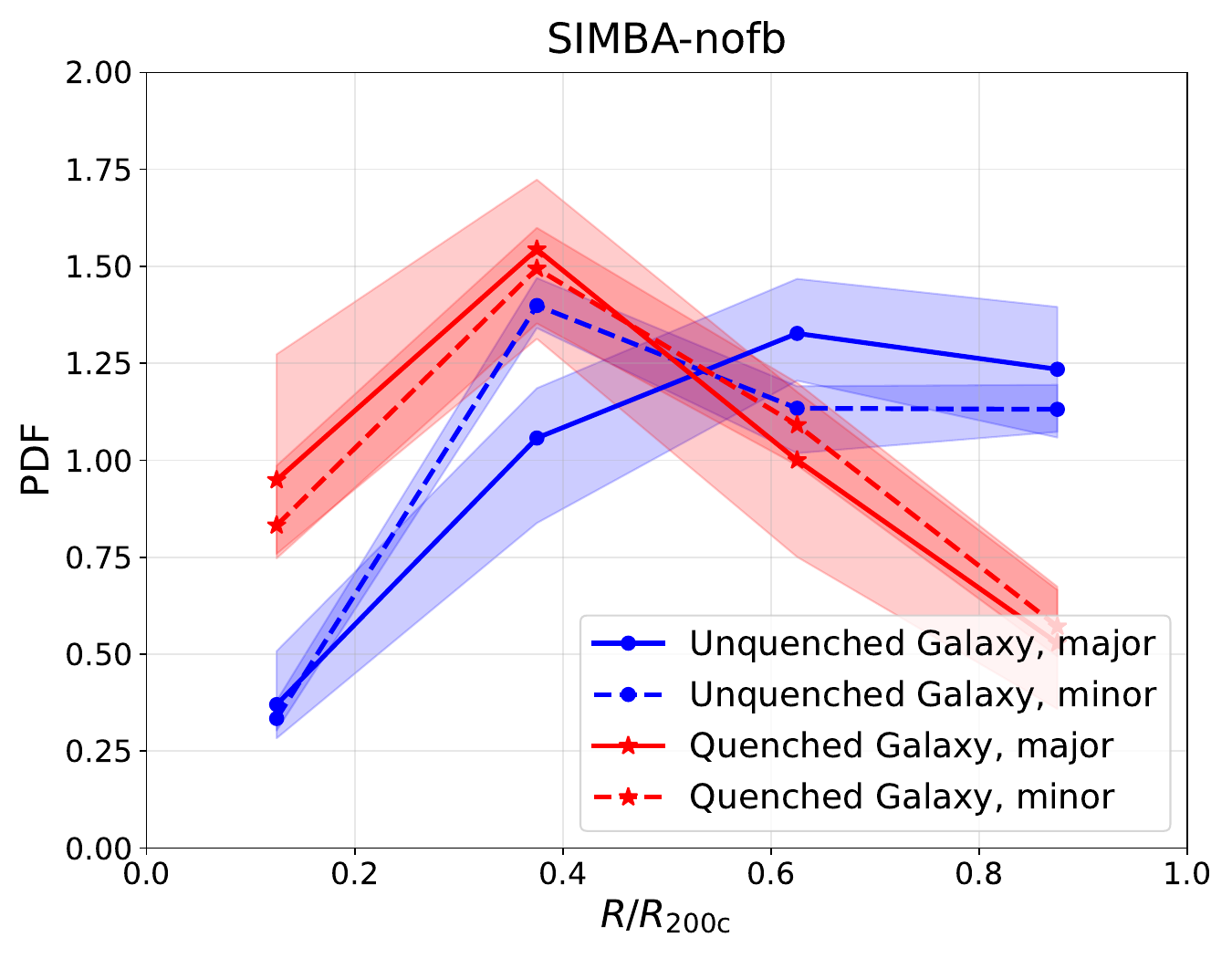}
\caption{
\textbf{Radial segregation of quenched and unquenched satellites.} Radial probability-density distributions of quenched and unquenched satellites with $M_{\star}\geq10^{9.5}\,\mathrm{M}_{\odot}$, separated by major and minor axes of the central galaxy. The four panels correspond to SIMBA, TNG100, EAGLE and SIMBA-nofb. Shaded regions show the $3\sigma$ confidence interval. Within $R_{200\mathrm{c}}$, quenched satellites are more centrally concentrated than unquenched satellites in all four simulations.}
\label{fig:UnQ_Q_R_BCG}
\end{figure}

To investigate the physical origin of the spatial anisotropy of quenched and unquenched satellite galaxies, we compare the radial distributions of the two populations in Figure~\ref{fig:UnQ_Q_R_BCG}, by showing their radial probability-density distributions of satellites with $M_{\star}\geq10^{9.5}\,\mathrm{M}_{\odot}$ in both major and minor axes. In all four simulations, quenched satellites within $R_{200\mathrm{c}}$ are more centrally concentrated, peaking at around $0.3\times R_{200c}$, than unquenched satellites along both major and minor axes, although the detailed profile shapes differ among the simulations. For the unquenched population, the radial probability-density distributions tend to rise or flatten with increasing radius rather than decline. This behaviour is consistently seen across the simulations and indicates a pronounced difference between the radial distributions of unquenched and quenched satellites. 

This radial segregation provides a physical connection between satellite quenching and the scale dependence of satellite anisotropy. In our companion study, satellites near the central galaxy were found to be more strongly aligned with the central stellar principal axes, with this alignment weakening at larger radii \citep{2026arXiv260427845Z}. The centrally concentrated quenched population therefore represents the satellite distribution aligned with the major axis of the central galaxy. By contrast, the more extended unquenched population is weighted towards larger radii, where the satellite distribution is less strongly correlated with the central stellar orientation. When host systems are stacked in the reference frame of their central galaxies, these differing radial weightings produce different directional number ratios. The quenched population retains a strong major-axis excess, whereas the unquenched population is closer to isotropic. Consequently,
\begin{equation}
\frac{N_{\mathrm{Q,major}}}{N_{\mathrm{Q,minor}}} > \frac{N_{\mathrm{U,major}}}{N_{\mathrm{U,minor}}},
\end{equation}
and a positive $\Delta F_{\mathrm{Q}}$ emerges. Crucially, quenching need not depend explicitly on azimuthal angle for this mechanism to operate. An angular dependence of the quenched fraction can arise from two independently measurable properties: the different radial distributions of quenched and unquenched satellites, and the intrinsic anisotropy distribution of the satellites. The key condition, established previously \citep{2026arXiv260427845Z}, is that the central galaxy influences the satellite distribution only within the inner region. That work examined this condition for SIMBA, TNG100 and EAGLE, but did not consider SIMBA-nofb; the condition therefore holds in three of our four simulations. Consequently, the ASGQ signals in SIMBA, TNG100 and EAGLE are expected, and the SIMBA-nofb signal indicates that the same mechanism may also hold in this no-feedback run. Although stellar and AGN feedback are not strictly required to generate this angular signal, the feedback-free simulation nonetheless contains a substantially different overall quenched population from the full-physics runs. This comparison therefore supports a feedback-independent origin for the geometrical ASGQ mechanism, but it does not imply that feedback is unimportant for determining which satellites become quenched. On the contrary, \cite{Cui2021} reported that the jet-mode feedback in SIMBA is the key to quenching galaxies. At radii beyond $R_{200\mathrm{c}}$, the difference between the quenched and unquenched directional ratios decreases and the radial ASGQ profiles become less consistent among these simulations. This transition is expected if the central stellar orientation becomes a weaker tracer of the surrounding galaxy distribution outside the host halo, where the anisotropic distribution of neighbouring groups, clusters and filaments becomes increasingly important, or if the baryon models play a more important role in the quenching fractions. We examine that regime in Section~\ref{sec:extra_origins}.

\subsection{Resolution effects seen from TNG and EAGLE}
\begin{figure}[h!]
\centering
\includegraphics[width=0.47\textwidth]{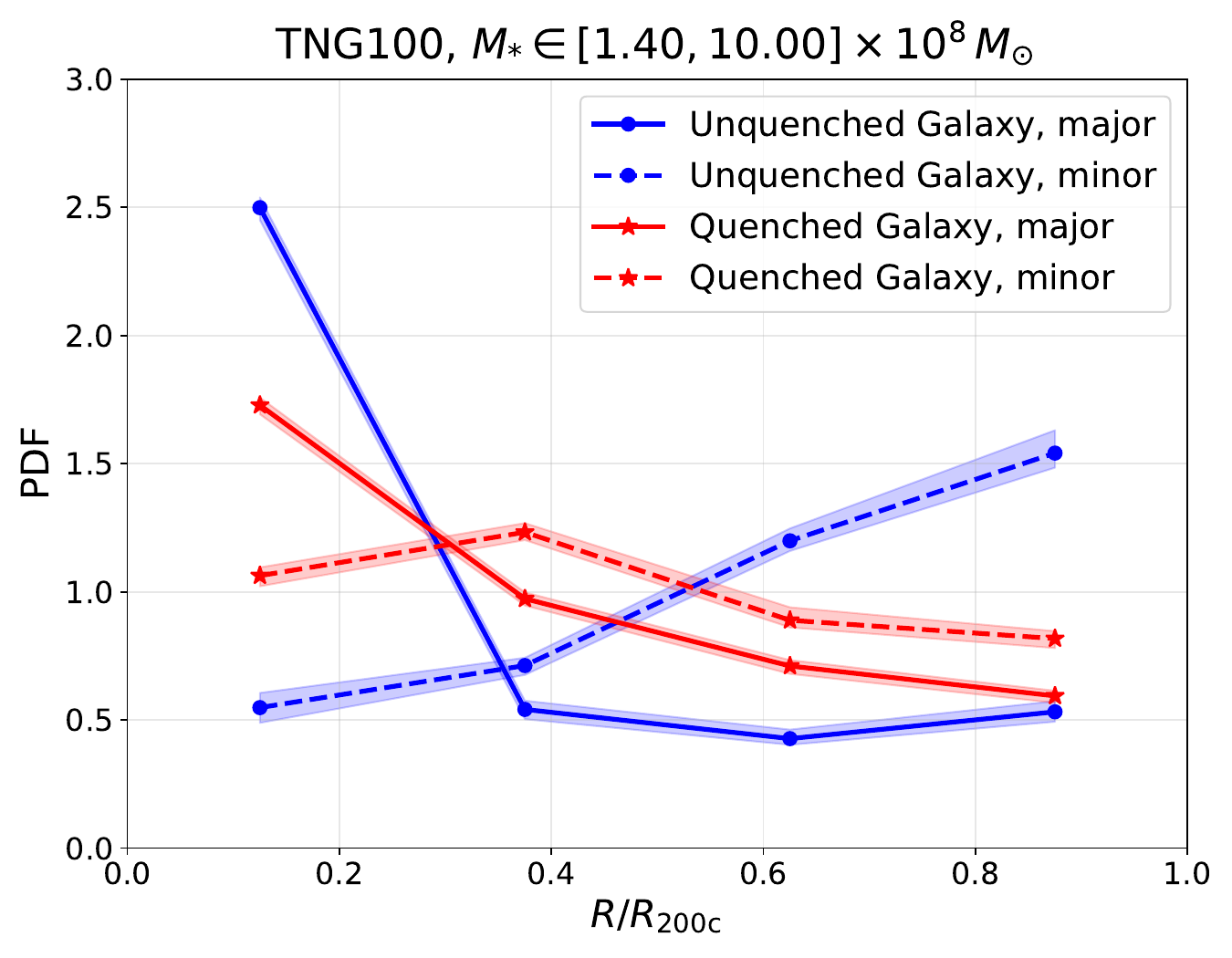}
\hspace{0pt}
\includegraphics[width=0.47\textwidth]{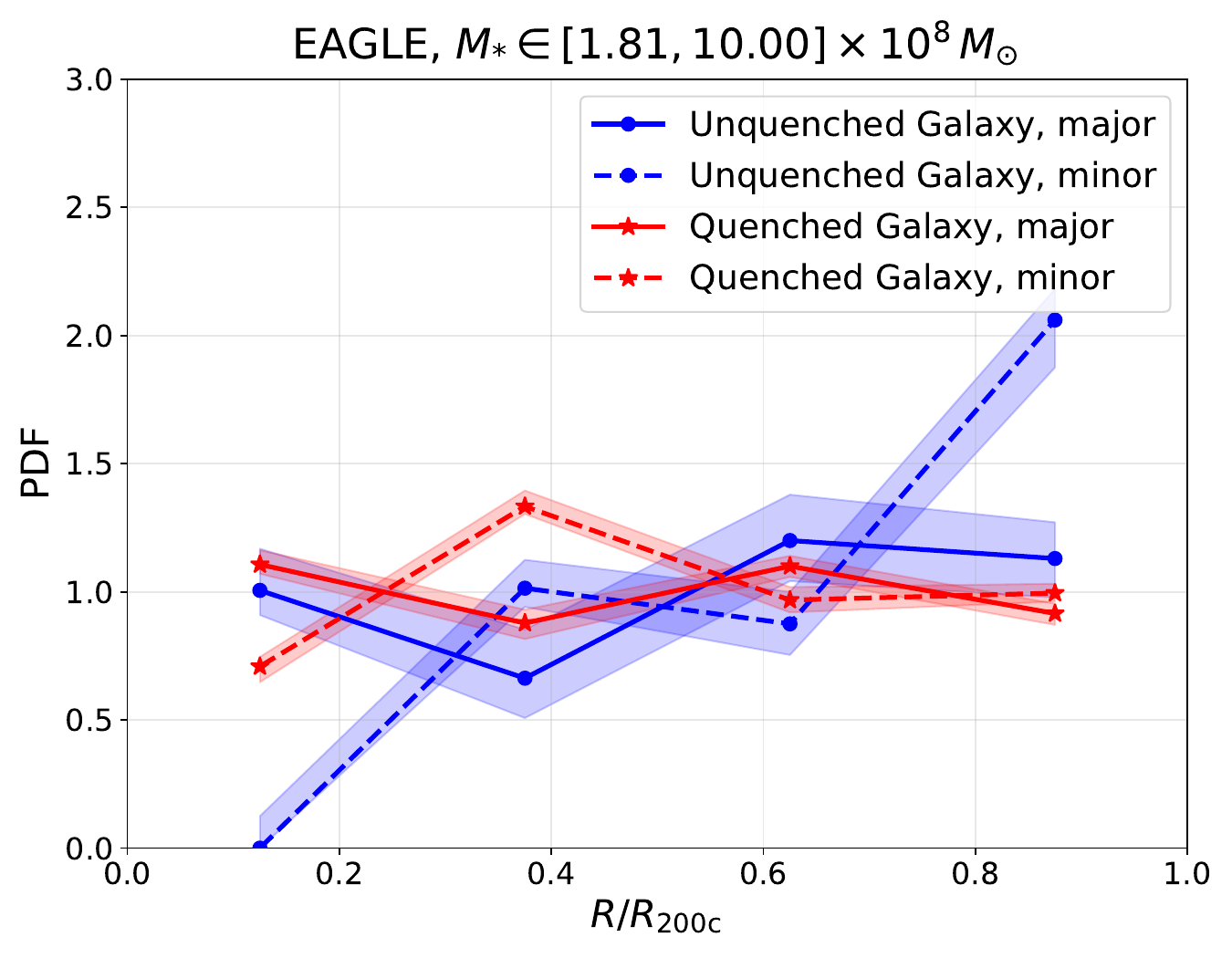}

\caption{
\textbf{Low-mass consistency test of the radial-segregation interpretation.} Radial probability-density distributions of quenched and unquenched satellites with $M_{\star}\leq10^{9}\,\mathrm{M}_{\odot}$ for TNG100 and EAGLE. The samples are separated according to their location along the major or minor axis of the central galaxy. Shaded regions show the $3\sigma$ confidence interval described in Methods. Unlike the primary massive samples, the low-mass samples do not exhibit a consistent central concentration of quenched satellites relative to unquenched satellites along the major axis.
}
\label{fig:lowmass_UnQ_Q_R_BCG}
\end{figure}

Our primary interpretation of the intra-halo ASGQ signal uses satellites with $M_{\star}\geq10^{9.5}\,\mathrm{M}_{\odot}$. This mass lies above the adopted threshold of $100\,m_{\mathrm{gas}}$ in all four simulations. The higher baryonic mass resolution of TNG100 and EAGLE also permits a supplementary test using satellites with $M_{\star}\leq10^{9}\,\mathrm{M}_{\odot}$. Figure~\ref{fig:ASGQ_Mstar_R} shows no clear intra-halo ASGQ signal in these low-mass TNG100 and EAGLE samples. We therefore use them as a consistency test of the radial-segregation interpretation. If a positive intra-halo ASGQ signal arises because quenched satellites are more centrally concentrated than unquenched satellites, then this radial ordering should be absent—or even reversed—in samples that do not show such a signal. Figure~\ref{fig:lowmass_UnQ_Q_R_BCG} shows the radial probability-density distributions of the low-mass quenched and unquenched populations. In the TNG100 major-axis selection, the unquenched population is at least as centrally concentrated as the quenched population across the relevant radial range. In EAGLE, the major-axis distributions of the two populations are similar, with no clear central excess of quenched satellites. The radial configuration associated with the massive-satellite ASGQ signal is therefore not reproduced in either low-mass sample. 

This comparison is consistent with the proposed connection between radial segregation and ASGQ. However, it is not used as an independent physical detection because the low-mass samples are more sensitive to numerical resolution, subhalo identification and the sampling of low star-formation rates. A quantitative interpretation of the low-mass dependence would require matched-volume, multi-resolution convergence tests. We therefore treat Figure~\ref{fig:lowmass_UnQ_Q_R_BCG} as a methodological robustness check rather than as one of the primary scientific results.

\section{The Physical Origins of ASGQ outside halos}\label{sec:extra_origins}

\begin{figure}[h!]
\centering
\includegraphics[width=0.47\textwidth]{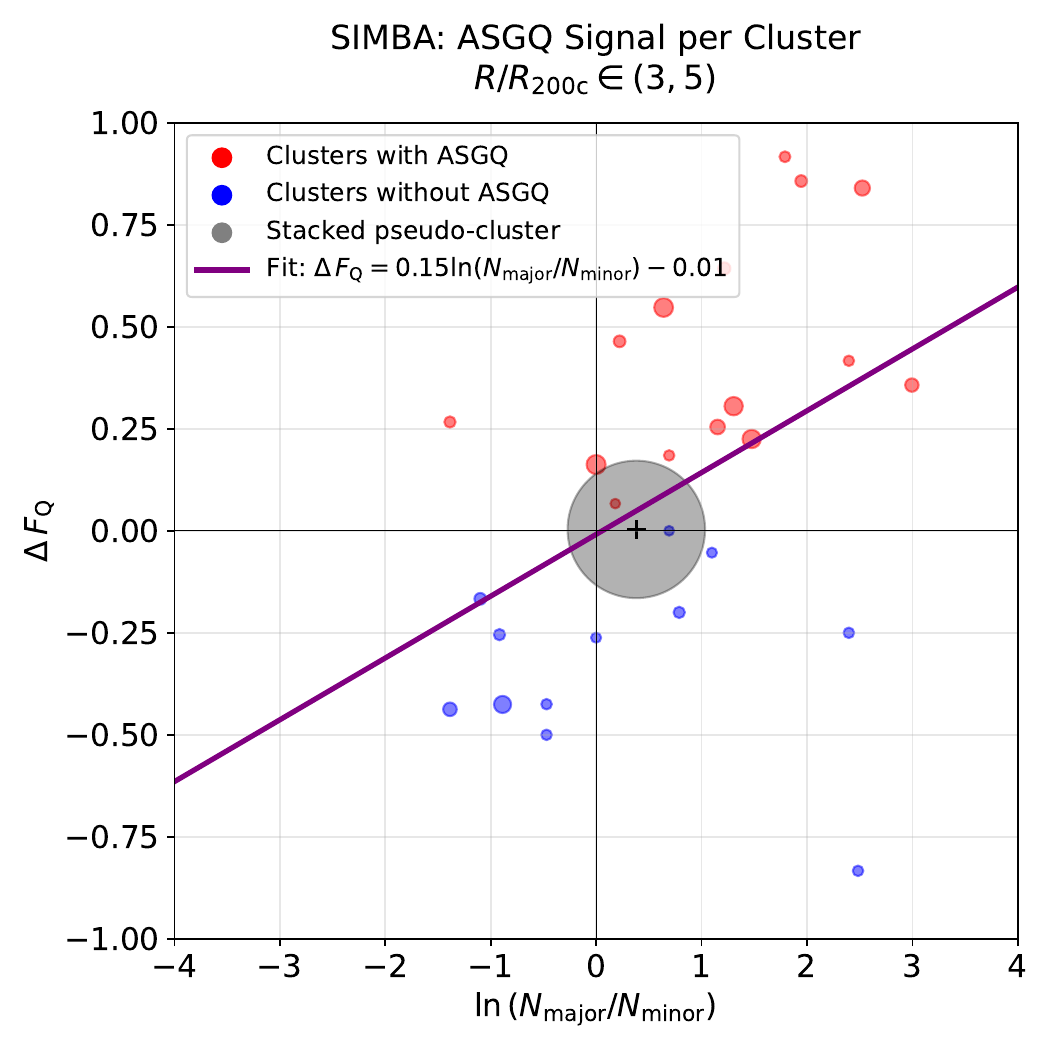}
\hspace{0pt}
\includegraphics[width=0.47\textwidth]{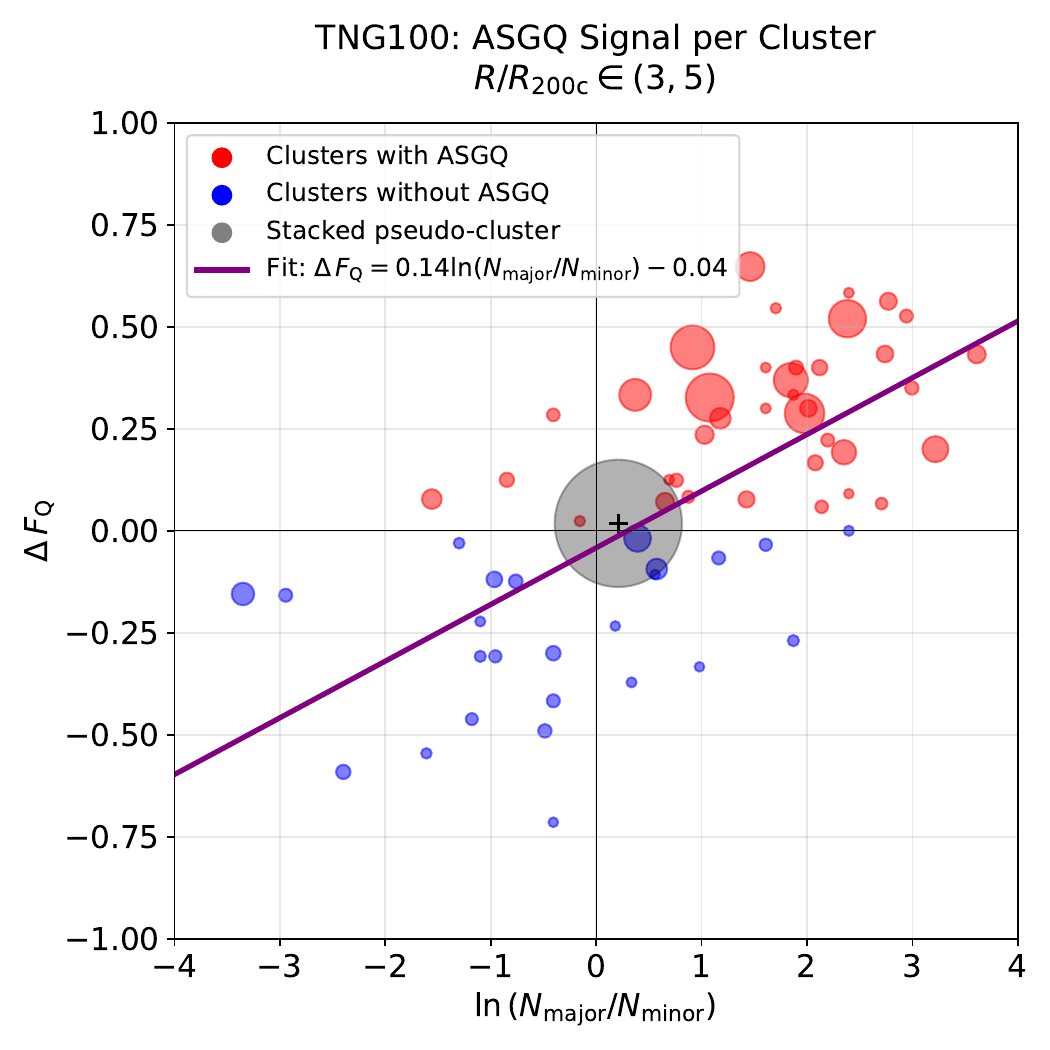}
\vspace{7pt}
\includegraphics[width=0.47\textwidth]{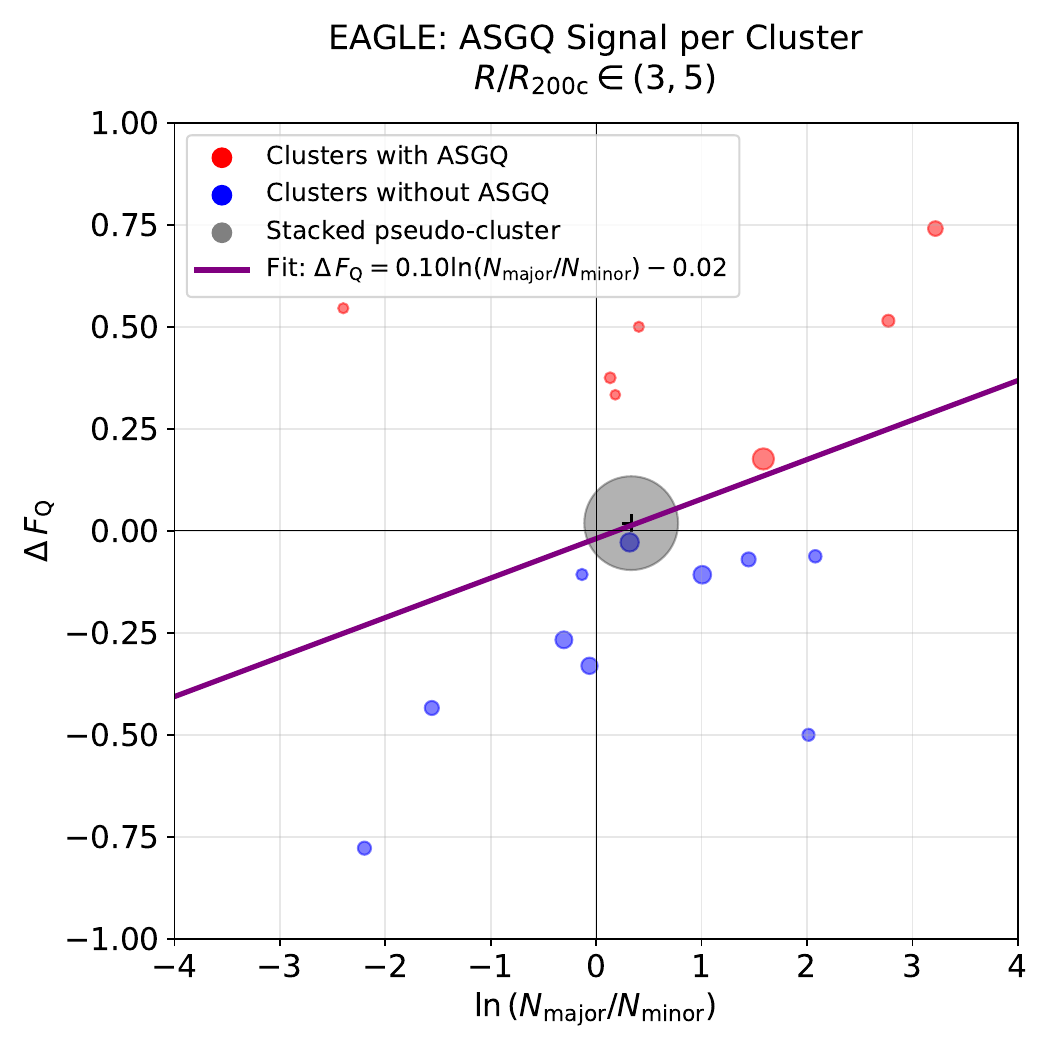}
\hspace{0pt}
\includegraphics[width=0.47\textwidth]{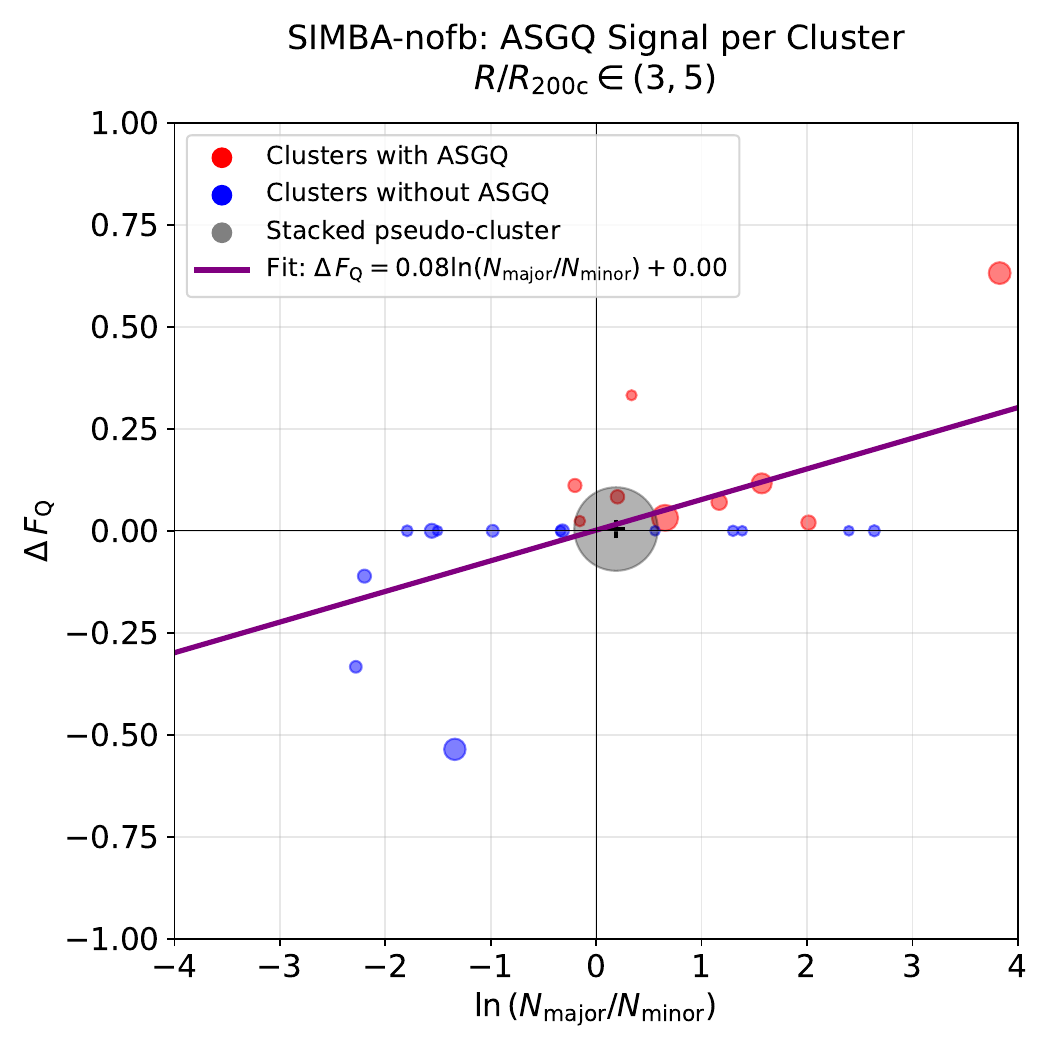}

\caption{
\textbf{Extra-halo ASGQ and directional galaxy-number asymmetry.} The quenched-fraction difference $\Delta F_{\mathrm{Q}}$ is shown against the logarithmic major-to-minor galaxy-number ratio for galaxies within $3<R/R_{200\mathrm{c}}<5$. The panels correspond to SIMBA, TNG100, EAGLE and SIMBA-nofb. Individual systems are shown when the combined major- and minor-axis galaxy count exceeds ten; marker size represents this combined number count. Systems below this count threshold are combined into a stacked pseudo-halo system, shown separately in a gray circle. The solid line is fitted to the individual systems only.  Across all simulations, individual halos exhibit a statistically significant positive correlation between ASGQ strength and axial satellite count ratio, while the pseudo-halos show weak or non-detectable ASGQ signal.}
\label{fig:ASGQ_Nratio}
\end{figure}

Outside the host-halo boundary, the spatial distribution of surrounding galaxies is less connected with the halo shape, as well as the morphology of the central galaxy. Our companion study found that satellite anisotropy at these distances is more closely associated with the filamentary environment \citep{2026arXiv260427845Z}. We therefore examine whether extra-halo ASGQ is related to directional variations by focusing on galaxies within $3<R/R_{200\mathrm{c}}<5$. This is because (1) this region is far from the influences of dark matter halos, and (2) there are larger differences between different simulations. For each simulation, we define a mass threshold of $M_{*} \ge 100\,m_\mathrm{gas}$, where $m_\mathrm{gas}$ is the gas particle mass as defined in Section \ref{sec:data}. Systems with more than ten galaxies satisfying this threshold in the combined major- and minor-axis selections are analysed individually. To explore the conditions under which halos produce a large-scale ASGQ signal and also remove the contamination from these low-statistics objects, of which the directional quenched fraction would be determined by only a few galaxies, systems with ten or fewer selected galaxies are combined after aligning the principal axes of their central galaxies, producing one stacked low-count sample for each simulation. 

Figure~\ref{fig:ASGQ_Nratio} compares $\Delta F_{\mathrm{Q}}$ with the major-to-minor galaxy-number ratio. In all four simulations, individual systems with positive $\Delta F_{\mathrm{Q}}$ preferentially occupy the region with $N_{\mathrm{major}}>N_{\mathrm{minor}}$. Systems with non-positive $\Delta F_{\mathrm{Q}}$ are more commonly found at $N_{\mathrm{major}}\leq N_{\mathrm{minor}}$. The fitted relations therefore have positive slopes, although the scatter and the number of individually resolved systems differ among the simulations. The stacked low-count samples have weak ASGQ amplitudes and lie very close to the fitting relations from the individually analysed systems. This behaviour indicates that the large-scale directional quenched fraction is also tightly connected with the galaxies' anisotropy distribution. However, we may not be able to see such ASGQ signal in all simulations, depending on the central galaxy's major and minor axes intersecting different large-scale environments. If the major axis points towards a galaxy-rich filament, group or neighbouring cluster, the corresponding direction contains more galaxies and produces $N_{\mathrm{major}}>N_{\mathrm{minor}}$. Galaxies in these dense structures may also have experienced stronger environmental processing or pre-processing than galaxies in less dense directions \citep{2018MNRAS.475.3654Z,2025MNRAS.537.1542S, 2025A&A...693A.113Z}. The same direction can consequently contain both a larger galaxy population and a larger quenched fraction. This interpretation is consistent with previous work connecting large-scale ASGQ to anisotropic accretion and filamentary pre-processing \citep{2025MNRAS.537.1542S,2025A&A...693A.113Z}. However, the correlation alone does not establish that a particular filament or neighbouring halo causes the measured quenching difference. Direct identification of the cosmic-web environment and neighbouring bound structures would be required to demonstrate that causal connection. To this end, analyzing galaxy clusters individually provides a more sensitive approach to uncovering the correlation between the ASGQ signal and galaxy-number asymmetry. In contrast, estimating the extra-halo ASGQ signal via stacking introduces a critical statistical limitation: both $\Delta F_{\mathrm{Q}}$ and $N_{\mathrm{major}}/N_{\mathrm{minor}}$ are derived from noisy and overlapping directional counts. These compound uncertainties and fluctuations tend to smear out the ASGQ signal, as shown in Figure \ref{fig:QF_R_BCG}. Therefore, we can interpret the observed relation as evidence for an association between extra-halo ASGQ and environmental anisotropy, rather than as a complete causal demonstration.

\section{Discussion}\label{sec:Discussion} 

Our comparison of SIMBA, TNG100, EAGLE and SIMBA-nofb identifies two scale-dependent contributions to anisotropic satellite quenching, both of which can originate from the anisotropic galaxy distribution. Within $R_{200\mathrm{c}}$, the four simulations consistently produce a positive ASGQ signal for satellites with $M_{\star}\geq10^{9.5}\,\mathrm{M}_{\odot}$. The signal is associated with a larger major-to-minor directional count ratio for quenched satellites than for unquenched satellites. This difference arises from radial segregation: the quenched population is more centrally concentrated and therefore samples more strongly the major-axis-aligned inner satellite distribution. The unquenched population is more radially extended and isotropic, distributed homogeneously with less association to the central stellar orientation. This mechanism does not require the quenching probability at a fixed radius to depend explicitly on the azimuthal angle. Instead, an apparent angular dependence of the quenched fraction emerges from combining a radial variation in satellite anisotropy with different radial distributions for quenched and unquenched galaxies. The appearance of the same qualitative intra-halo behaviour in SIMBA-nofb has a specific implication. Stellar feedback, AGN feedback and X-ray heating are not necessary for producing the geometrical ASGQ mechanism identified here. However, this does not imply that feedback is unimportant for satellite quenching. The substantially lower massive-galaxy quenched fraction in SIMBA-nofb shows that feedback strongly modifies the overall galaxy population. Therefore, it may affect the amplitude of the ASGQ signal as well as its stellar-mass dependence. 

Outside the halo, the simulations show greater variation. TNG100 exhibits a particularly strong large-scale contribution from low-mass galaxies, whereas SIMBA, EAGLE and SIMBA-nofb generally produce weaker extra-halo signals. On a system-by-system basis, $\Delta F_{\mathrm{Q}}$ is positively associated with the directional galaxy-number ratio. This relation suggests that extra-halo ASGQ is more physically connected to the anisotropic distribution of surrounding groups, clusters and filaments. Directions containing richer galaxy environments may also contain a larger fraction of galaxies that have experienced environmental processing or pre-processing. The higher ASGQ signal in TNG100 reveals its higher possibility of alignment between its central galaxies' orientation and the large-scale filament outside the halo, which may deeply connected with its baryon and hydrodynamic models. 

The intra-halo and extra-halo results can therefore be understood within a single scale-dependent geometrical framework. Inside the halo, the relevant effect is the radial segregation of quenched and unquenched satellites combined with the radial dependence of satellite alignment. Outside the halo, the relevant effect is the direction-dependent abundance and environmental history of galaxies in the surrounding large-scale structure. The two regimes are connected through hierarchical assembly but need not have the same immediate physical origin. 

Several limitations should be considered. The simulations differ in volume, mass resolution, hydrodynamical method and baryonic implementation. Differences in their low-mass and extra-halo results cannot therefore be attributed uniquely to feedback. Our satellite definition is also geometrical and includes galaxies that are not gravitationally bound to the central halo, particularly outside $R_{200\mathrm{c}}$. This broad definition is useful for studying the surrounding environment but differs from the usual bound-subhalo definition of a satellite. Another missing or lack-of-statistics part in this study is the limited number of galaxy clusters, which is restricted by the simulation volumes, generally $\lesssim 150$ Mpc. The ASGQ signal is also reported in galaxy cluster by \cite{2022MNRAS.511.2659S}. In addition, the present analysis uses three-dimensional simulated stellar principal axes, whereas observational measurements normally use projected positions and photometric position angles. Projection, interloper contamination and central-axis measurement errors should be included in a direct comparison with observations. Finally, the interpretation of the extra-halo relation requires explicit identification of cosmic filaments, neighbouring haloes and pre-processed galaxy populations.

\section{Conclusions} \label{sec:conclusions}
In this study, we provide a cross-simulation assessment of the scale-dependent ASGQ effect across SIMBA, TNG100, EAGLE, and SIMBA-nofb. Synthesizing these numerical results, our main conclusions are structured as follows:

\begin{itemize}
    \item \textbf{Robust Intra-halo Signal:} All four simulations consistently show a positive intra-halo ($<R_{200\mathrm{c}}$) ASGQ signal for massive satellites ($M_{\star} \geq 10^{9.5}\,\mathrm{M}_{\odot}$).\\
    \item \textbf{Intra-halo ASGQ via Satellite Radial Segregation:} Inside $R_{200\mathrm{c}}$, quenched satellites are centrally concentrated with a preference along the central galaxy's major axis, whereas unquenched satellites are more radially extended and isotropic. Combining these distinct radial profiles produces an apparent azimuthal quenching preference without requiring explicit angle-dependent quenching physics.\\
    \item \textbf{Extra-halo ASGQ via Large-Scale Structural Asymmetry:} Outside $R_{200\mathrm{c}}$, extra-halo ASGQ positively tracks the major-to-minor galaxy count ratio. This connects the large-scale signal to the directional asymmetry of the surrounding galaxy population, pointing to the underlying large-scale environmental structure.\\
    \item \textbf{Role and Impact of Feedback:} The qualitative intra-halo signal in SIMBA-nofb confirms feedback is unnecessary for its generation, though feedback might modulate the total quenched fraction, signal amplitude, and stellar-mass dependence.
\end{itemize}

Future observational tests should jointly measure the radial distributions of quenched and star-forming satellites, their directional number ratios and the angular distribution of galaxies beyond the halo boundary. Such measurements would distinguish the spatial-segregation mechanism proposed here from a direction-dependent change in quenching probability at fixed radius and would test whether the extra-halo relation is present in observed groups and clusters. Furthermore, we are also planning to extend this study to galaxy clusters, which have the most abundant satellites, with a larger sample \citep[using e.g.][]{Cui2018}, as well as to high-redshift, to understand this particular signal from dynamical and evolutionary perspectives.

\begin{acknowledgments}
Z. Z. and Y. C. would like to thank Dr. Liang Gao for his helpful discussions. This work has been supported by the National Key Research and Development Program of China (No.\ 2022YFA1602903), the National Natural Science Foundation of China (Nos.\ 12588202 and 12473002), and the China Manned Space Program with grant no.\ CMS-CSST-2025-A03. 
W.C. gratefully thanks Comunidad de Madrid for the Atracci\'{o}n de Talento fellowship no. 2020-T1/TIC19882 and Agencia Estatal de Investigaci\'{o}n (AEI) for the Consolidaci\'{o}n Investigadora Grant CNS2024-154838. He further acknowledges the Project PID2024-156100NB-C21, financed by MICIU/AEI/10.13039/501100011033/FEDER, and the science research grants from the China Manned Space Project.
\end{acknowledgments}

\vspace{5mm}

\bibliography{new}{}
\bibliographystyle{aasjournal}

\end{document}